\documentclass[useAMS,usenatbib]{mn2e}
\usepackage{amsmath}
\usepackage{amssymb}
\newcommand{\gdet}{\sqrt{-g}}

\usepackage{graphics}
\usepackage{graphicx}
\usepackage{epstopdf}
\usepackage{footnote}
\usepackage{tablefootnote}
\usepackage{txfonts}
\usepackage{lipsum}
\usepackage{float}
\usepackage[draft]{hyperref}
\usepackage{macros}
\usepackage{fixltx2e}

\makesavenoteenv{tabular}

\DeclareSymbolFont{cmletters}{OML}{cmm}{m}{it}
\DeclareMathSymbol{v}{\mathalpha}{cmletters}{"76}

\usepackage[usenames,dvipsnames,svgnames,table]{xcolor}
\usepackage{hyperref}
\definecolor{darkblue}{rgb}{0.0,0.0,0.3}
\hypersetup{colorlinks,breaklinks,
            linkcolor=darkblue,urlcolor=darkblue,
            anchorcolor=darkblue,citecolor=darkblue}

\title[]{Magnetically Modified Spherical Accretion in GRMHD: Reconnection-Driven Convection and Jet Propagation
 }
\author[S. M. Ressler, E. Quataert, C. J. White, O. Blaes ]{S. M. Ressler$^{1},$  E. Quataert$^{2,3},$ C. J. White$^{3}$, O. Blaes$^{4}$\\
$^{1}$Kavli Institute for Theoretical Physics, University of California Santa Barbara, Santa Barbara, CA 93107 \\
$^{2}$Departments of Astronomy \& Physics, Theoretical Astrophysics Center, University of California, Berkeley, CA 94720 \\
$^{3}$Department of Astrophysical Sciences, Princeton University, Princeton, NJ 08544 \\
$^{4}$Department of Physics ,University of California, Santa Barbara, CA 93106, USA}

\begin{document}

\maketitle

\begin{abstract}
We present 3D general relativistic magnetohydrodynamic(GRMHD) simulations of zero angular momentum accretion around a rapidly rotating black hole, modified by the presence of initially uniform magnetic fields.  
We consider serveral angles between the magnetic field direction and the black hole spin.
In the resulting flows, the midplane dynamics are governed by magnetic reconnection-driven turbulence in a magnetically arrested (or a nearly arrested) state. 
Electromagnetic jets with outflow efficiencies $\sim$ 10--200\% occupy the polar regions, reaching several hundred gravitational radii before they dissipate due to the kink instability.
The jet directions fluctuate in time and can be tilted by as much as $\sim$ 30$^\circ$ with respect to black hole spin, but this tilt does not depend strongly on the tilt of the initial magnetic field. 
A jet forms even when there is no initial net vertical magnetic flux since turbulent, horizon-scale fluctuations can generate a net vertical field locally.
Peak jet power is obtained for an initial magnetic field tilted by 40--80$^\circ$ with respect to the black hole spin because this maximizes the amount of magnetic flux that can reach the black hole.
These simulations may be a reasonable model for low luminosity black hole accretion flows such as Sgr A* or M87.
  \end{abstract}

\section{Introduction}

 Well-resolved, three dimensional general relativistic magnetohydrodynamic (GRMHD) simulations of accreting black holes are now commonplace \citep{DeVilliers2003,Gammie2003,White2016,BHAC,HAMR} and are frequently used to model low luminosity accretion flows such as those surrounding the supermassive black holes Sagittarius A* (Sgr A*) and the one at the center of M87 \citep{Mosci2009,Mosci2016,CK2015,EHT5,Chael2019,Dexter2020}.  The most common strategy is to initialize the simulations with a torus in hydrodynamic equilibrium (e.g., \citealt{Fishbone1976,Penna2013}) and add a dynamically unimportant magnetic field to seed the magnetorotational instability (MRI, \citealt{BalbusHawley}) and drive accretion.   By varying the size of the torus, the initial magnetic field geometry,  and the magnitude of the black hole spin, a relatively diverse set of solutions emerge that can, in principle, be used to constrain properties of observed systems.  In order to do so precisely, one must understand not only the dependence of the results on these particular parameters, but also on the assumptions that go into each model.  

For instance, much recent work has been devoted to exploring the effects of non-ideal physics expected to be important for low luminosity accretion flows.  This includes two-temperature effects \citep{Ressler2015,Sadowski2017,Chael2018}, non-thermal electron acceleration 
\citep{Mao2016,Ball2016,Chael2017}, radiation \citep{BHLIGHT,Sadowski2017,Ryan2017,Chael2019}, magnetic resistivity \citep{Ripperda2019}, and anisotropic conduction/viscosity \citep{grim,Foucart2017}.  These works build on earlier 1D semi-analytic models that more easily incorporated extra physics (e.g., \citealt{Narayan1995,Ozel2000,Yuan2003,Johnson2007}). Another active research frontier is that of tilted accretion disks, where the spin of the black hole is misaligned with the angular momentum of the flow \citep{Fragile2007,Liska2018,Liska2019,White2019b,Liska2020}.   Tilted disks not only have different dynamics than non-tilted disks (e.g., precession, standing shocks), they can also produce qualitatively different images/spectra (e.g. ellipsoidal morphologies, larger image size, \citealt{White2020,Chatterjee2020}).  

Less explored are the effects of the torus initial conditions.   This was the motivation for a series of multi-scale simulations of Sgr A* (\citealt{Ressler2018,Ressler2020,Ressler2020b}, building on earlier work by \citealt{Cuadra2005,Cuadra2006,Cuadra2008}, see also \citealt{Calderon2020}) that used observationally-constrained Wolf-Rayet stellar wind source terms in order to essentially eliminate the need for assumed initial conditions.  Two of the key findings of these studies were that the gas ultimately reaching the horizon had: 1) a relatively small amount of angular momentum, and 2) a relatively large amount of coherent magnetic flux. As a consequence, the MRI had little effect on the dynamics and the flow became magnetically arrested \citep{Narayan2003,Igumenshchev2003} on horizon scales (at least for $a = 0$, where $a$ is the dimensionless spin of the black hole).   Although we do not typically have such intricate knowledge of the stellar/gas population surrounding supermassive black holes in other galaxies, it is not unreasonable to assume that some fraction of these objects are fed in a similar way. 

In this work we seek to investigate an alternative model of low angular momentum, magnetized accretion flows in GRMHD.    In particular, we consider the limiting case of a non-rotating flow at large radii with an initially uniform magnetic field, the MHD analog of the classic force-free \citet{Wald1974} and \citet{Bicak1985} solutions.
In non-relativistic simulations \citep{CDBF,Pen2003,Pang2011}, such flows have been shown to have distinct characteristics from those with significant angular momentum, e.g., turbulence driven by magnetic reconnection and nearly hydrostatic force balance.  For a non-spinning black hole, the qualitative nature of these solutions likely carries over to full GRMHD,  though with order unity quantitative corrections near the horizon.  Conversely, for a rapidly spinning black hole, the large supply of coherent magnetic flux will presumably drive a \citet{BZ1977} jet that would back-react on the quasi-spherical inflow and could significantly alter the resulting dynamics.  Additionally, frame dragging and magnetic torque could lead to non-negligible azimuthal velocities near the horizon even if there is initially no rotation.

One potentially important parameter in these simulations is the angle that the initial magnetic field makes with the black hole spin axis.  Previous works that study tilt have used equilibrium tori seeded with various magnetic field geometries contained within those tori (e.g., \citealt{Fragile2007,Liska2018,Liska2019,White2019b,Liska2020}).   This means that the orientation of the magnetic field was tilted along with the angular momentum of the gas, making it hard to disentangle the effects of magnetic tilt vs. angular momentum tilt.  Here, starting with no angular momentum, we can investigate the magnetic tilt in isolation in order to determine how much, if any, it effects the orientation of both the jet and the accretion flow. In particular, we focus on four magnetic tilt angles, $0^\circ,30^\circ, 60^\circ$, and $90^\circ$, with an initial plasma $\beta$ of 100, a Bondi radius of 200 gravitational radii, and a dimensionless black hole spin $a=0.9375$.  

This work proceeds as follows.  \S \ref{sec:model} outlines the initial conditions and numerical model, \S \ref{sec:results} presents the resulting simulations, \S \ref{sec:60} analyzes further one particular simulation that is an outlier from the rest, \S \ref{sec:param} discusses the dependence of our results on various parameters, \S \ref{sec:comp} compares these results to previous work, and \S \ref{sec:conc} concludes. 


\section{Model}
\label{sec:model}

Since non-radiative GRMHD simulations are scale-free, we will generally use length and time units that scale with the mass of the black hole and set the speed of light, $c$, and the gravitational constant, $G$, equal to unity.    Thus, in these units,  the mass of the black hole, $M$, is our standard unit for both length and time, though we often use the gravitational radii $r_{\rm g} = M$ as an equivalent notation.

Our simulations are carried out in {\tt Athena++} \citep{White2016,Athena++}, a 3 dimensional grid-based code that solves the equations of conservative GRMHD in Cartesian Kerr--Schild coordinates (CKS, \citealt{Kerr1963}).  These relate to the spherical Kerr--Schild $r, \theta, \varphi$ coordinates via 
\begin{equation}
    \label{eq:xyz}
\begin{aligned} 
  &x = r \sin(\theta) \cos(\varphi) + a\sin(\theta)\sin(\varphi) \\
 &y = r \sin(\theta) \sin(\varphi) - a \sin(\theta)\cos(\varphi) \\
 &z = r \cos(\theta), 
 \end{aligned}
\end{equation}
where $a$ is the dimensionless spin of the black hole.   Our computational domain is a rectangular box of size $3200 r_{\rm g}\times 3200 r_{\rm g}\times 25600 r_{\rm g}$ with base resolution $24 \times 24 \times 192$ cells in the $x,y,$ and $z$ directions.  The box is larger in the $z$-direction in order to ensure that any relativistic jet that might form does not interact with the outer boundary.    We use an additional 12 levels of nested static mesh refinement (SMR) to achieve approximately logarithmic spacing in $r$, using meshblocks of size $12 \times 24 \times 16$ cells.  The outer boundary of the first level is located at $|z| = 6400 r_{\rm g}$ and $|x|,|y| = 1600 r_{\rm g}$, while the outer boundary of the second level is located at $|z| = 3200 r_{\rm g}$ and $|x|,|y| = 1600 r_{\rm g}$.  The outer boundaries for the remaining $n \ge 3$ levels are located at  $|x|,|y|, |z|= 1600/2^{n-3} r_{\rm g}$.   The finest grid spacing is achieved for $|x|,|y|, |z|< 3.125 r_{\rm g}$, where $\Delta x,\Delta y,\Delta z \approx 0.0326 r_{\rm g}$, ensuring that the event horizon is well resolved since $(\Delta x,\Delta y,\Delta z)/r_{\rm H} \ll 1$, where $r_{\rm H} = 1 + \sqrt{1-a^2}$ is the event horizon radius.  All of our simulations use $a = 0.9375$, for which $r_{\rm H} \approx 1.35 r_{\rm g}$. Note that for $|x|,|y|, |z|\le 1600 r_{\rm g}$, the grid is equivalent to a $(1600 r_{\rm g})^3$ box with a $192^3$ base resolution with 9 levels of extra mesh refinement.  Our resolution is comparable to or better than the highest resolution ($192^3$) spherical KS torus simulations presented in the EHT code comparison project that were found to be converged  \citep{Porth2019}.  Specifically, at $(r=12 r_{\rm g},\theta={\rm \pi}/2)$, our simulations are better resolved by almost a factor of 2 in the radial direction and more than a factor of 2.5 in the $\varphi$ direction, while having a comparable resolution in the $\theta$ direction to the most midplane-refined of their simulations.

The initial conditions for these simulations are inspired by previous non-relativistic models of magnetically modified spherical accretion in the literature (e.g., \citealt{CDBF,Pen2003,Pang2011}).  Outside of $r = 6 r_{\rm g}$, we start with a uniform density, $\rho_0 =1$, and uniform pressure, $P_0$, such that the Bondi radius is $r_{\rm B} \equiv 2 M/c_{s,0}^2 = 200 M$, where $c_{s,0} = \sqrt{\gamma P_0/\rho_0} $ is the non-relativistic, adiabatic sound speed.  These non-relativistic expressions are appropriate for large Bondi radii where GR effects are small.   For simplicity, the gas is stationary, with the spatial components of the four-velocity set to zero.  Inside $r = 6 r_{\rm g}$, the density and pressure are set to the numerical floors and again the spatial components of the four-velocity are set to zero.  We set the initial magnetic field by taking the curl of a vector potential $A_\mu$,
\begin{equation}
  B^i = \frac{1}{\sqrt{-g}}\epsilon^{ijk} \frac{\partial }{\partial x^j} A_k,
  \label{eq:curlA}
\end{equation}
where $g=-1$ is the determinant of the CKS metric and $\epsilon^{ijk}$ is the Levi--Cevita symbol.   All spatial components of $A_\mu$ are zero except for the $y$-component (note that the $t$-component is irrelevant), which  we set to be $A_y \propto e^{5 (1-6/r)} \left[-z \sin(\psi) + x \cos(\psi)\right] $, with the normalization set such that $\beta \equiv P/P_{B} = 100 $ at large radii, where $P_{B} =b^2/2$ is the magnetic pressure, $b^2 = b^\mu b_\mu$, and $b^\mu$ is the magnetic field four vector in Lorentz--Heaviside units.  This vector potential corresponds to a uniform field that is inclined by an angle of $\psi$ with respect to the black hole spin axis for $r > 6 r_{\rm g}$.   For $r \lesssim 6 r_{\rm g}$ the field exponentially decays to zero, meaning that the field lines wrap around $r \approx 6$ as shown in the upper left panels of Figures \ref{fig:Bz_contour}--\ref{fig:Bx_contour}.

Inside $r<r_{\rm H}/2$, we set the density/pressure to the numerical floors and the four-velocity to free fall.  Within this region the induction equation is solved without alteration.  Although CKS coordinates are horizon-penetrating, there is still a coordinate singularity at $z=0$, $\sqrt{x^2+y^2}=|a|$ (i.e., $r=0$, $\theta=\pi/2$).  To avoid this, for $|z|<10^{-5} r_{\rm g}$ we set $z=10^{-5} r_{\rm g}$ in the calculation of the metric.  Since this is well inside the event horizon, it has no effect on our results.
The outer boundary conditions at the box faces are fixed to the initial conditions.

The density floor is $10^{-6} (r/r_{\rm g})^{-3/2}$ and the pressure floor is $3.33 \times 10^{-9} (r/r_{\rm g})^{-5/2}$, with $\sigma \equiv b^2/\rho \le 100 = \sigma_{\rm max}$ and $\beta \ge 0.001$ enforced via additional density and pressure floors, respectively.  Here $\beta$ is the ratio between the thermal and magnetic pressures.  Additionally, the velocity of the gas is limited such that the maximum Lorentz factor is 50 in the CKS normal frame. 

We run simulations with $\psi = 0^\circ,30^\circ,60^\circ,$ and 90$^\circ$ for 20,000 $M$, about 10 free-fall times at the Bondi radius.\footnote{The $\psi =90^\circ$ simulation uses a slightly different grid than the other three simulations in that the $|z|>1600 r_{\rm g}$ region is excluded.  Within $|x|,|y|,|z|<1600 r_{\rm g}$ the grids are identical. }  We set the adiabatic index to $\gamma=5/3$.

Finally, we note that for analysis purposes we will often convert the simulation data from CKS coordinates $x,y$, and $z$ to spherical KS coordinates $r, \theta$, and $\varphi$ using the inverse of Equations \eqref{eq:xyz}.  Four-vectors are converted using the Jacobian:
\begin{equation}
  A^\mu_{(\rm KS)} = \frac{\partial x^\mu_{(\rm KS)}}{\partial x^\nu_{(\rm CKS)}} A^\nu_{(\rm CKS)}.
\end{equation}
We will also make use of an orthonomal tetrad to measure local velocities in the zero angular momentum (ZAMO) frame, which we define in Boyer--Lindquist coordinates \citep{BL} as an observer with four-velocity
\begin{equation}
  \eta_\mu^{\rm BL} = \left(-\frac{1}{\sqrt{-g^{tt}_{\rm BL}}},0,0,0\right).
\end{equation}
The corresponding tetrad is
\begin{equation}
  \begin{aligned}
  e^\mu_{t,\rm BL} &= \eta^\mu_{\rm BL} \\
  e^\mu_{r,\rm BL} &= \left(0,\frac{1}{\sqrt{g_{rr}^{\rm BL}}},0,0\right)\\
  e^\mu_{\theta,\rm BL} &=\left(0,0,\frac{1}{\sqrt{g_{\theta \theta}^{\rm BL}}},0\right) \\
  e^\mu_{\phi,\rm BL} &=\left(0,0,0,\frac{1}{\sqrt{g_{\varphi \varphi}^{\rm BL}}}\right),  \\
  \end{aligned}
\end{equation}
so that the physical velocities in the $ith$ direction are 
\begin{equation}
  V^i = \frac{u^\mu_{\rm KS} e_\mu^{i,\rm KS}}{u^\mu_{\rm KS} e _\mu^{t,\rm KS}},
\end{equation}
where $e^\mu_{\nu,\rm KS}$ is $e^\mu_{\nu,\rm BL}$ converted to KS coordinates.

\section{Results}
\label{sec:results}
\subsection{Overview}
\label{sec:overview}
\begin{figure*}
\includegraphics[width=0.95\textwidth]{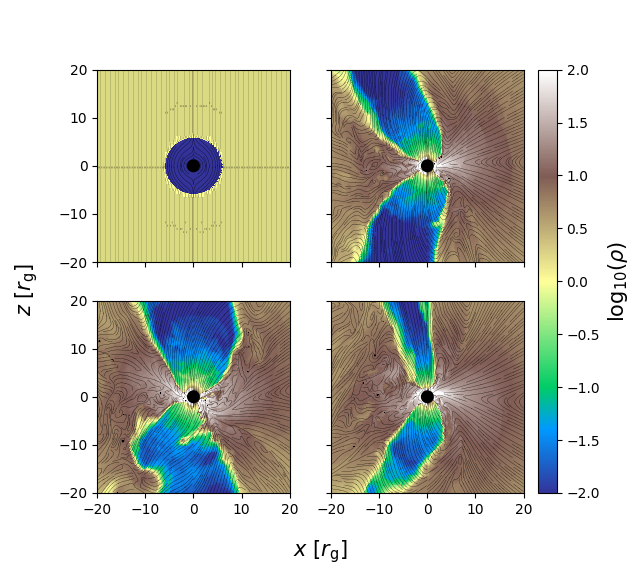}
\caption{2D slice through the $x$--$z$ plane of mass density over-plotted with magnetic field lines at four different times in the $\psi = 0^\circ$ simulation.  Starting from the top left panel and proceeding clockwise, the panels represent 0, 4000, 10000, and 18000 $M$.  The initially uniform field accretes through the event horizon and the rapidly spinning black hole drives a powerful jet.  The accreting field also reconnects, heating the gas and driving turbulence.  An animated version of this figure is available at \href{https://youtu.be/N7wz-vPLwLo}{https://youtu.be/N7wz-vPLwLo}. } 
\label{fig:Bz_contour}
\end{figure*}

\begin{figure*}
\includegraphics[width=0.95\textwidth]{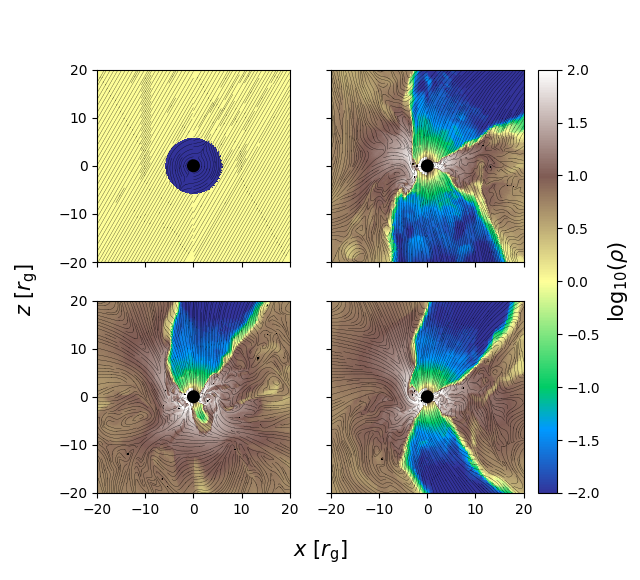}
\caption{Same as Figure \ref{fig:Bz_contour} but for $\psi = 30^\circ$.} 
\label{fig:30_contour}
\end{figure*}

\begin{figure*}
\includegraphics[width=0.95\textwidth]{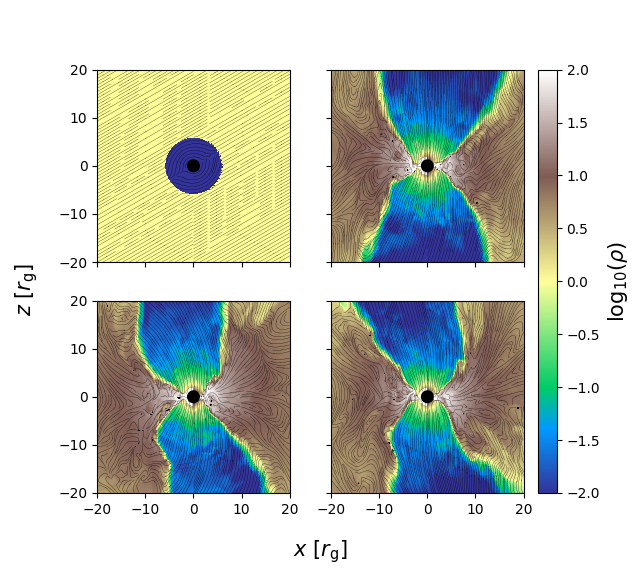}
\caption{Same as Figure \ref{fig:Bz_contour} but for $\psi = 60^\circ$.} 
\label{fig:60_contour}

\end{figure*}

\begin{figure*}
\includegraphics[width=0.95\textwidth]{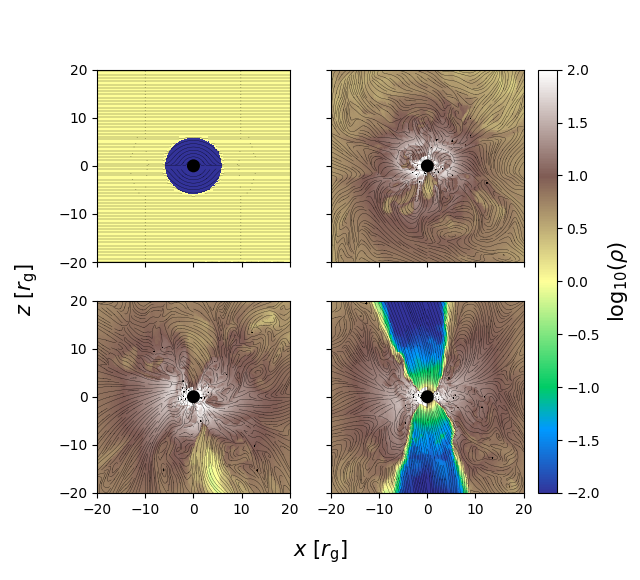}
\caption{Same as Figure \ref{fig:Bz_contour} but for $\psi = 90^\circ$. At 4000 $M$ (top right panel), the gas is in a state of disordered convection that persists until a net vertical magnetic field is randomly generated locally and a jet is formed.  An animated version of this figure is available at \href{https://youtu.be/L3apzf2cTP4}{https://youtu.be/L3apzf2cTP4}. } 
\label{fig:Bx_contour}
\end{figure*}

Figures \ref{fig:Bz_contour}--\ref{fig:Bx_contour} show 2D contour slices of the mass density over-plotted with magnetic field lines at four different times for our $\psi = 0^\circ$, $30^\circ$, $60^\circ$, and $90^\circ$ simulations, respectively.  In all cases, the initially weak magnetic field is dragged inwards with the flow until it becomes dynamically important.  Absent black hole spin, this would result in a `pinched' geometry of the field where matter is pushed away from the magnetic poles and towards the magnetic midplane.  In non-relativistic MHD this ultimately leads to a highly disordered state where heating associated with the reconnection of the pinched field drives convection \citep{CDBF,Pen2003,Pang2011}.   Here, however, the effects of the rapidly spinning black hole significantly alter this scenario.    For $\psi = 0^\circ, 30^\circ$ and $60^\circ$, the accumulation of flux threading the horizon (particularly flux parallel to the black hole spin) caused by the initially spherical accretion leads to the formation of a magnetically dominated jet by $\sim$ 2000 $M$ via the \citet{BZ1977} mechanism in which the magnetic field is tightly wound up in the $\varphi$-direction and propels matter/energy outwards.  We find that the jet is always aligned with the black hole spin for $r\lesssim 5 r_{\rm g}$, though at larger radii it can be significantly tilted (we discuss jet orientation in more detail in \S \ref{sec:jet}).   The strong outflow evacuates the polar regions of matter but leaves a broad distribution of gas in the midplane. 
Turbulence driven by magnetic reconnection is visible at small ($r \lesssim 5 r_{\rm g}$) radii in Figures \ref{fig:Bz_contour}--\ref{fig:Bx_contour}.

The $\psi =90^\circ$ simulation shows similar behavior between 5000--15000 $M$ but much different behavior at earlier and later times.  Initially, the dynamics are governed by reconnection driven turbulence with no preferred direction, the GRMHD analog to the non-relativistic simulations mentioned above.  This state persists until turbulent fluctuations spontaneously generate a net locally vertical magnetic field that accretes and powers a jet.  However, since this vertical field is not being continuously supplied by gas at large radii, the jet dies off and the convection dominated state resumes.  We expect that if the simulation were to run for a longer time then another turbulently-generated, locally vertical field would be created and the cycle would repeat.  

\begin{figure*}
\includegraphics[width=0.95\textwidth]{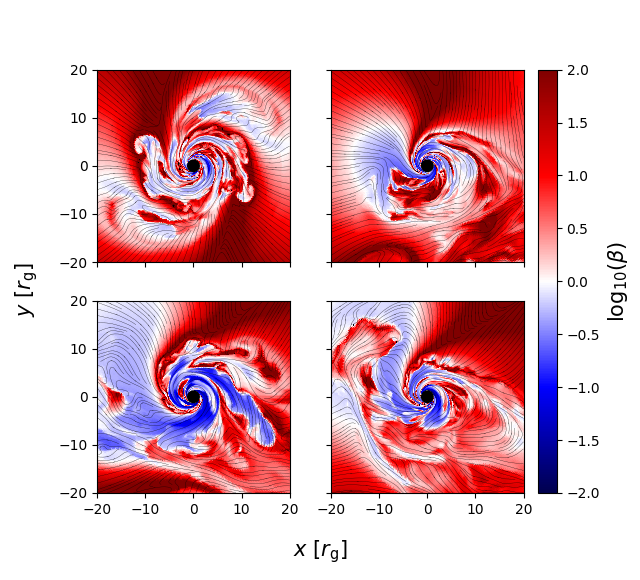}
\caption{Midplane slice of plasma $\beta$ at four different times in our $\psi = 30^\circ$ simulation overplotted with magnetic field lines.  Starting from the top left panel and proceeding clockwise, the panels represent 2000, 8000, 14000, and 20000 $M$.  Accretion proceeds along thin, high density streams surrounded by highly magnetized, low $\beta$ regions. Contours of $\beta$ in the other three simulations look similar.} 
\label{fig:beta_contour}
\end{figure*}

The simulations also have interesting azimuthal structure.  Figure \ref{fig:beta_contour} shows plasma $\beta$ in the midplane of the $\psi =30^\circ$ simulation as an example at four different times (the other three simulations display similar contours and are thus not shown). The flow structure is highly nonaxisymmetric, with clear spiral structure caused by the rotating spacetime of the central Kerr black hole.  Such structures are largest and most coherent in the $\psi=90^\circ$ simulation since the twisting of the initially horizontal field by the black hole provides the most spatially extended torque on the gas.  In all cases, matter generally accretes in thin streams with higher $\beta$ ($\gtrsim$ 1) regions surrounded by lower $\beta$ ($\lesssim$ 1) magnetically dominated regions.


%

\begin{figure}
\includegraphics[width=0.45\textwidth]{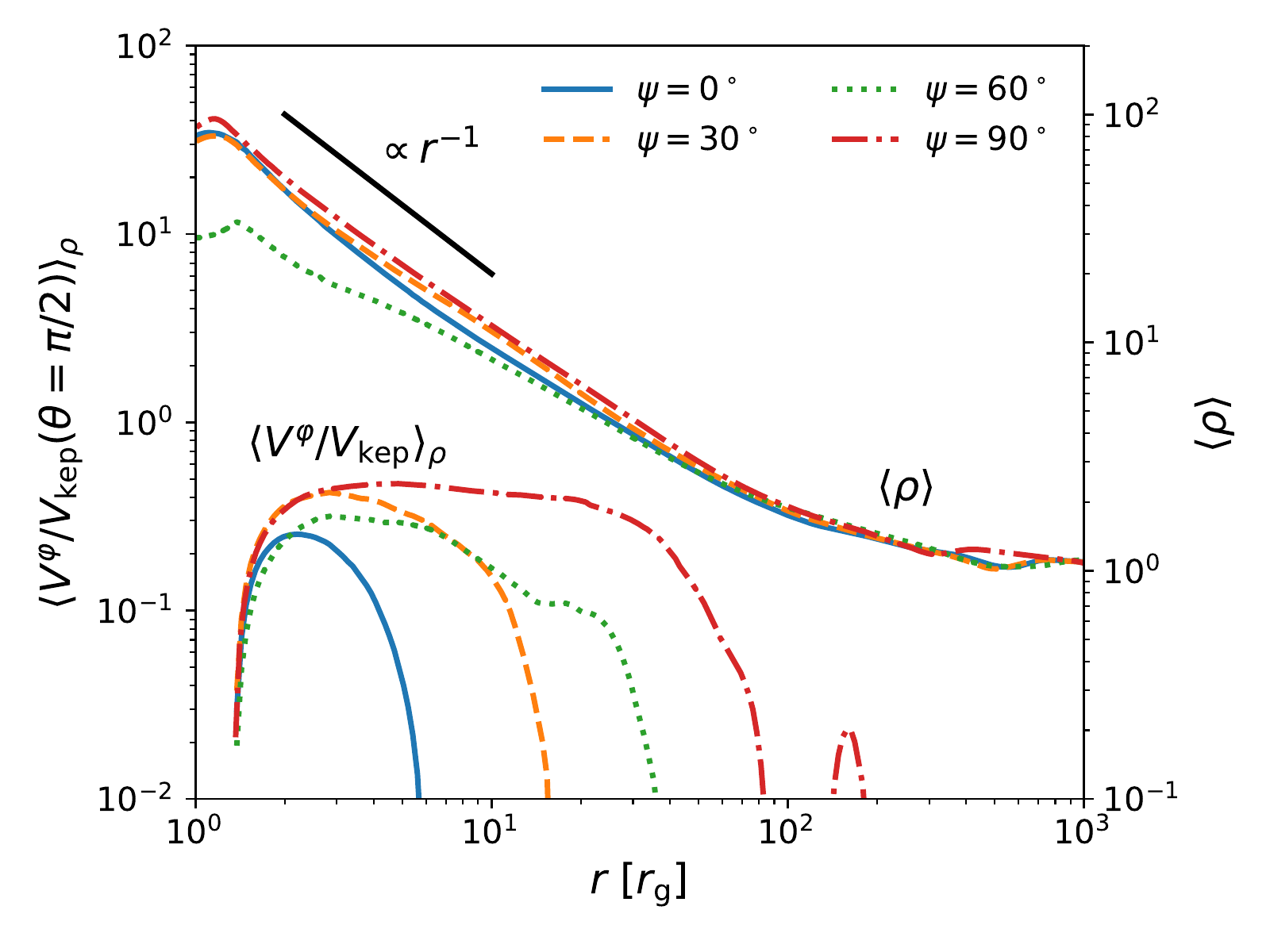}
\includegraphics[width=0.45\textwidth]{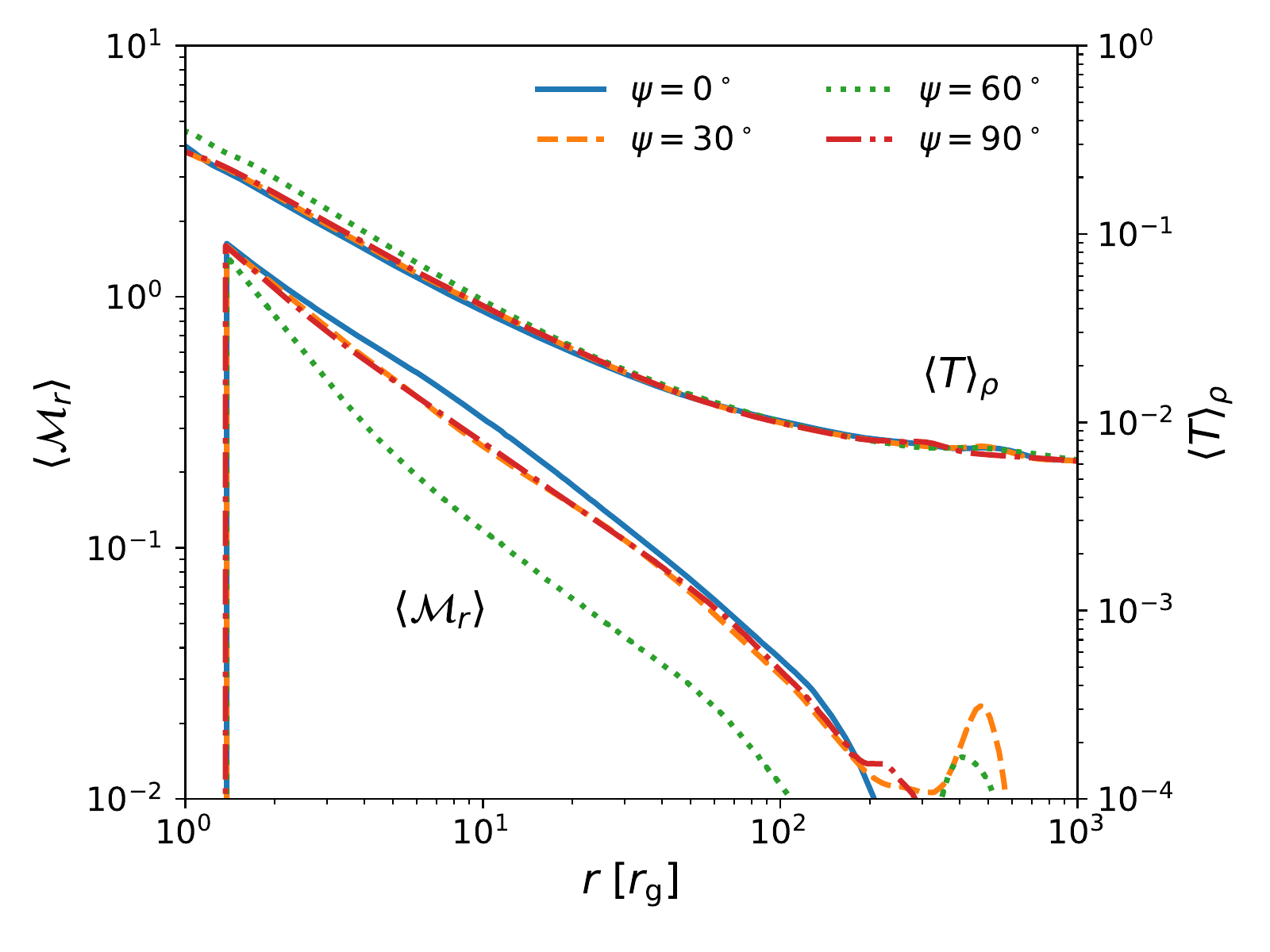}
\includegraphics[width=0.45\textwidth]{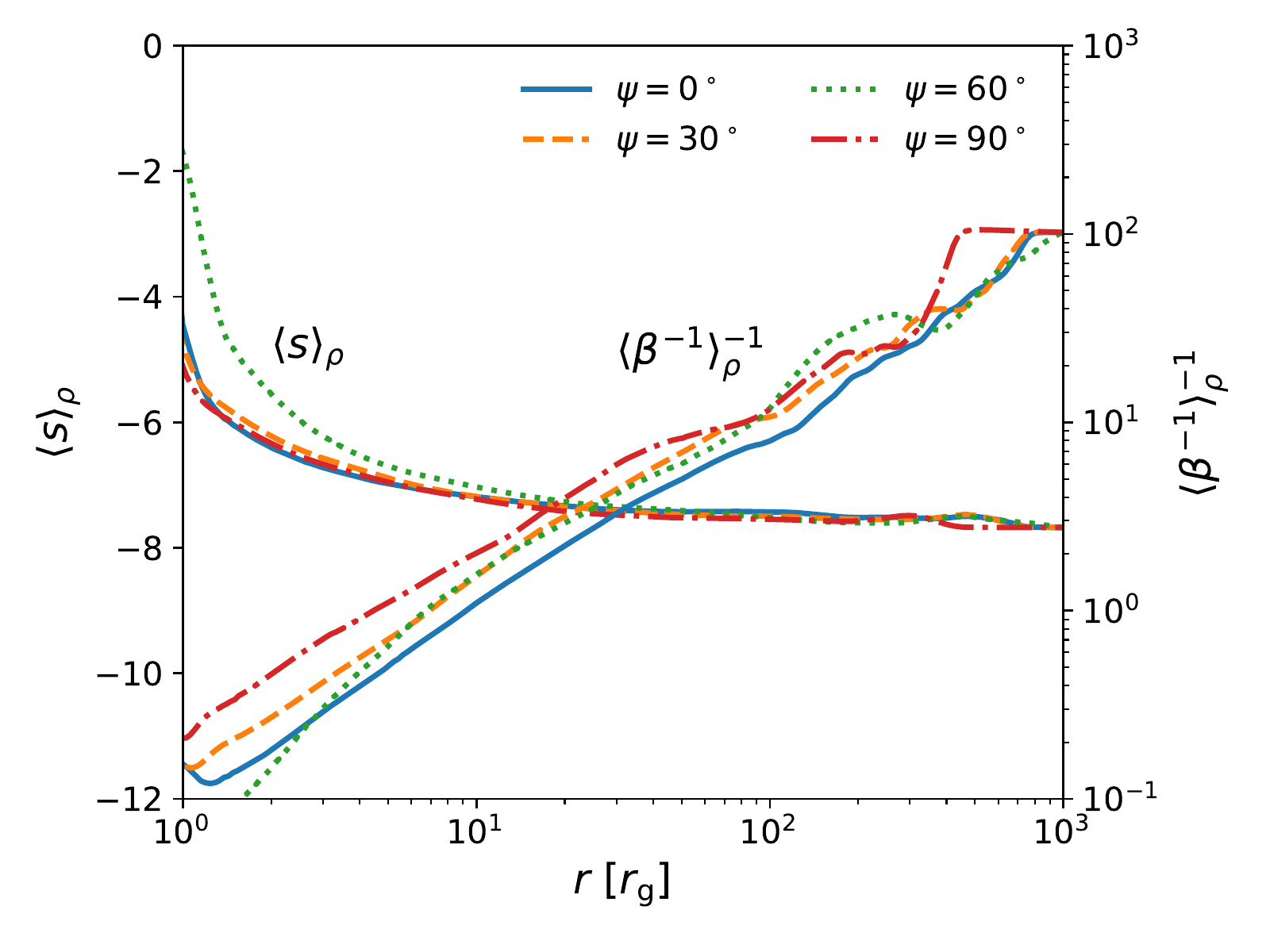}
\caption{Angle and time averaged fluid quantities plotted versus radius in each of our four simulations. Top: midplane angular velocity relative to Keplerian in the ZAMO frame, $\langle V_\varphi (\theta={\rm \pi}/2)\rangle _\rho/V_{\rm kep} $, and mass density, $\langle \rho \rangle$.  Middle: radial velocity divided by the sound speed in the ZAMO frame, $\langle\mathcal{M}_r \rangle_\rho$, and the temperature, $\langle T\rangle _\rho$.  Bottom: entropy, $\langle (\gamma-1)^{-1}\log (P  /\rho^\gamma)\rangle_\rho $, and plasma $\beta$,  $\langle \beta ^{-1}\rangle^{-1}_\rho$.  Note that $\log$ here represents the natural logarithm. The accretion flow in these simulations is transonic, but the sonic radii at $\sim$ $2r_{\rm g}$ is much closer to the event horizon than for the $a=0$ hydrodynamic Bondi solution, where the sonic radius is $\sim$ 9$r_{\rm g}$.  This is primarily due to the dynamically important, $\beta \lesssim 1$ magnetic fields located within $r\lesssim 10 r_{\rm g}$.  Torque exerted by these fields being dragged by the rotating Kerr spacetime also leads to significant ($\sim 0.5 V_{\rm kep}$) rotational velocities near the horizon $r\lesssim 10$--30 $r_{\rm g}$.  }
\label{fig:radial_plots}
\end{figure}

Figure \ref{fig:radial_plots} shows time and angle averages of several important quantities for the four simulations during the interval 10000--15000 $M$.  These include the mass density, $\rho$; the midplane angular velocity relative to the prograde Keplerian value measured in the ZAMO frame, $V^\varphi/V_{\rm kep}$, where $V_{\rm kep}$ is computed by transforming $\dot \varphi^{\rm BL}_{\rm Kep} = 1/(r^{3/2} + a)$ to the ZAMO frame; the radial velocity divided by the sound speed measured in the ZAMO frame, $V^r/c_{\rm s}$, where $c_{\rm s} = \sqrt{\gamma P/w}$ and $w = \rho + \gamma/(\gamma-1) P$; temperature, $T_{\rm g}  = P/\rho$ (assuming ionized hydrogen); entropy, $s= (\gamma-1)^{-1} \log(P/\rho^\gamma)$; and plasma $\beta$.  Here we compute angle averages in KS coordinates by evaluating
\begin{equation}
  \langle A\rangle_{X} =  \frac{\iint A X d\Omega }{\iint X d\Omega},
\end{equation}
where $d\Omega = \sqrt{-g_{\rm KS}} d\varphi d\theta$ and $g_{\rm KS} = -\sin(\theta)^2 [r^2 + a^2 \cos(\theta)^2 ]^2$.  The quantities plotted in Figure \ref{fig:radial_plots} are typically mass-weighted in order to study the properties of the accreting midplane.  We will discuss the properties of the jet later in \S \ref{sec:jet}.   Qualitatively, the four simulations display the same basic trends in the mass-weighted radial profiles.  Within $r\lesssim 10 r_{\rm g}$, the magnetic field has been compressed to the point that the flow is magnetically dominated, with $\beta \lesssim 1$, providing a Lorentz force strong enough to slow the infall to the degree that the sonic point is located at $\sim$ 2 $r_{\rm g}$ instead of the hydrodynamic Bondi value of $\sim$ 8.8 $r_{\rm g}$.\footnote{The hydrodynamic Bondi radius for $a=0$ is calculated analytically \citep{Michel1972,Hawley1984} .  We have verified in a 3D hydrodynamic simulation that a spherical flow onto an $a=0.9375$ black hole has the same sonic point radius.}  Note, however, that this Lorentz force is still $\lesssim$ the thermal pressure force because the magnetic pressure and tension terms tend to act in opposite directions, reducing the overall net magnetic force on the gas. The strong field is reconnecting and heating the gas, as evidenced by the entropy increasing with decreasing $r$ and the associated temperature increase beyond what one would predict from adiabatic compression alone.\footnote{It is somewhat surprising to see the entropy continue to increase after the fluid passes through the sonic point ($r \lesssim 2 r_{\rm g}$) and plunges through the event horizon at $r \approx$ 1.35 $r_{\rm g}$ because n this narrow radial range one would expect the gas to essentially be experiencing adiabatic free-fall.  This may be because 1) averaging effects result in an average radial entropy profile that is not necessarily representative of the entropy profile along individual streamlines or 2) the thermodynamics in this region may be less reliable than at larger radii because $\sigma \gtrsim 1$.}   Inside the Bondi radius (200 $r_{\rm g}$) the mass density settles into an $\rho$ $ \tilde \propto$ $ r^{-1}$ profile.
Finally, despite the stationary, spherical initial conditions, the frame-dragging effects of the $a=0.9375$ black hole on the magnetic field lines torque the gas enough to create non-neglible prograde rotational velocities for $r\lesssim 20$--50 $r_{\rm g}$, reaching up to $\sim$ 0.5 Keplerian near the horizon.  Note that this is indeed a torquing effect in the sense that $u_\varphi$ is nonzero and $u^\varphi$ is larger than what one would expect from frame dragging of the gas alone (in which case $V^\varphi$ would be zero). Compared to the gas in the $\psi = 0^\circ$ and $\psi = 30^\circ$ simulations, the gas in the $\psi = 60^\circ$ and $\psi = 90^\circ$ simulations has a shallower rotation curve for $r\gtrsim 3$ $r_{\rm g}$, meaning that significant angular velocity is present out to slightly larger radii.  This is likely because the larger amount of field perpendicular to the spin axis provides a longer `lever arm' to torque the flow.

The biggest outlier in Figure \ref{fig:radial_plots} is the $\psi = 60^\circ$ simulation, which has a $\sim $ 2--3 times slower radial velocity at all radii compared to the other simulations, as well as a $\sim $ 2--3 times lower density, a $\lesssim 2 $ times lower temperature, and a $\sim$ 2 times higher $P/\rho^\gamma$  for $r\lesssim$ 10 $r_{\rm g}$.  As we will show in the next subsection, this is because the amount of flux able to reach the black hole for $\psi = 60^\circ$ is larger than the other three simulations.  We speculate as to why in \S \ref{sec:60}.

\begin{figure*}
\includegraphics[width=0.95\textwidth]{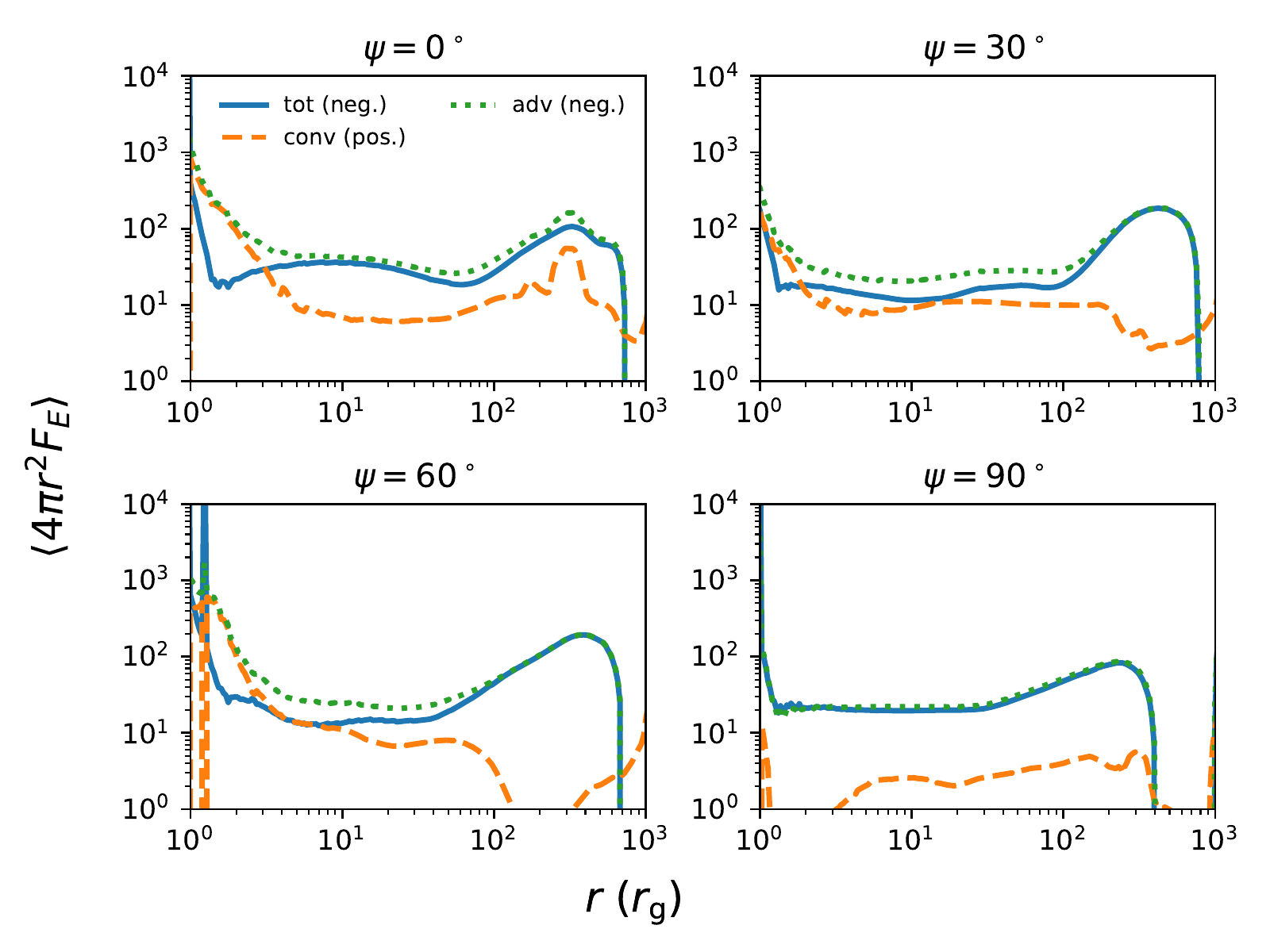}
\caption{Flux of energy in our four different simulations, averaged over time and angle broken down into the total (solid), convective (dashed), and advective (dotted) components.  Here we restrict the averages to $|\theta-\pi/2| < \theta_{H}$, where $\theta_{H}$ is the half-angle associated with the scale height of the disc.  Convection transports outwards only a fraction ($\lesssim 1/3$) of the energy being advected inwards except at the innermost radii ($r \lesssim$ 5--10 $r_{\rm g}$).   The convective flux is particularly small in the $\psi = 90^\circ$ simulation, $\lesssim$ 0.1 times the advective flux at all radii.}
\label{fig:F_conv_comp}
\end{figure*}

In terms of energy flux, we find that convection is inefficient at transporting energy outwards.  Figure \ref{fig:F_conv_comp} demonstrates this by plotting the average total flux of energy contained in matter,
\begin{equation}
  F_{\rm E, Ma}=-\frac{4 {\rm \pi}}{3}  \left(3r^2 + a^2\right)\left\langle \left(1 + \frac{\gamma}{\gamma-1} \frac{P}{\rho}\right)\rho u^r u_t  - \rho u^r\right\rangle,
  \end{equation}
 compared to the average flux of matter energy through laminar advection, 
 \begin{equation} 
   F_{\rm adv,Ma}= -\frac{4 {\rm \pi}}{3}  \left(3r^2 + a^2\right) \left\langle \left(1 + \frac{\gamma}{\gamma-1} \frac{P}{\rho}\right)u_t -1\right\rangle  \langle \rho u^r \rangle,
    \end{equation} 
    and the average convective flux, $F_{\rm conv,Ma} =F_{\rm E, Ma}- F_{\rm adv,Ma}$, for all four simulations.  
    In averaging these terms, we exclude the low density jet by focussing only on $|\theta - {\rm \pi}/2|< \theta_{H}$, where $\theta_{H} = \langle |\theta-{ \rm \pi}/2|\rangle_\rho$ is the half opening angle of the mass distribution.
    Energy flows inwards predominantly by laminar advection, while convection provides only a small outwards contribution with $|F_{\rm conv,Ma}|$ a factor of 3--10 times smaller than  $|F_{\rm adv,Ma}|$ except for $r\lesssim $ a few $r_{\rm g}$ where the two fluxes can be comparable.   The lack of a strong outwards convective flux explains why the density profile in Figure \ref{fig:radial_plots} is steeper than the $r^{-1/2} $ predicted by models where convection is the dominant energy transport mechanism (e.g., \citealt{Narayan2000,CDAF,CDBF}).

\subsection{Time Evolution}
\label{sec:time_evol}
\begin{figure}
\includegraphics[width=0.45\textwidth]{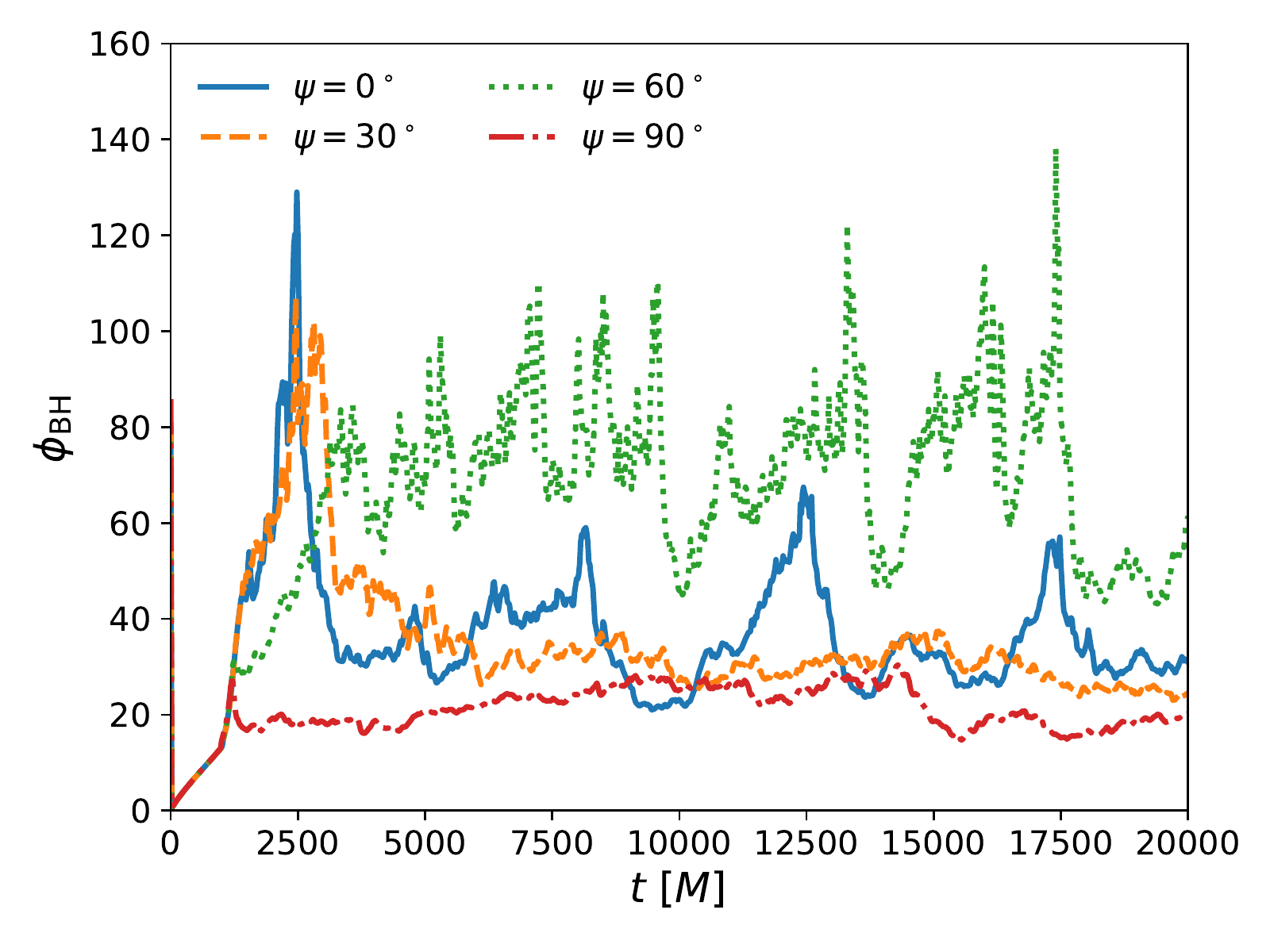}
\includegraphics[width=0.45\textwidth]{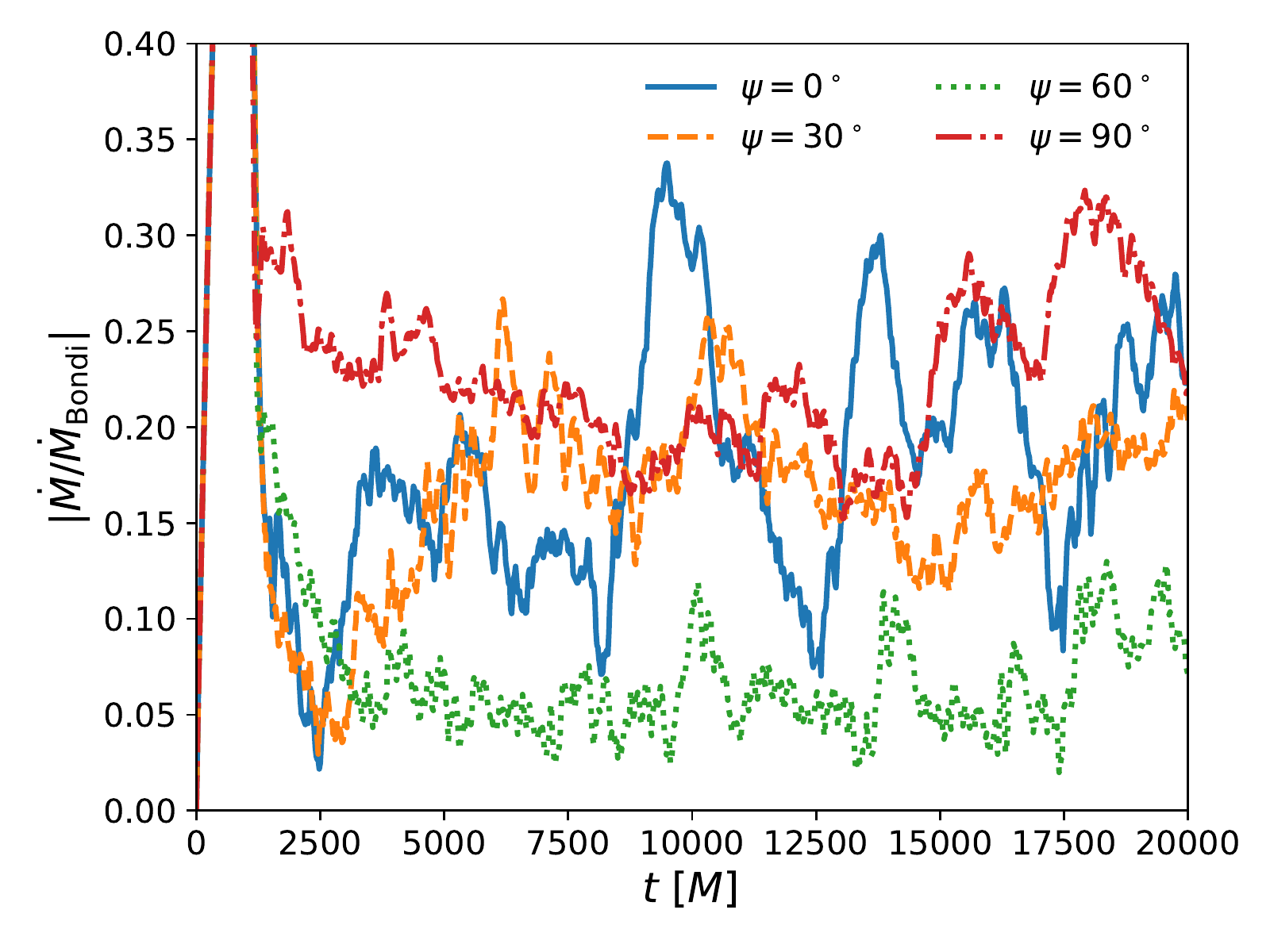}
\includegraphics[width=0.45\textwidth]{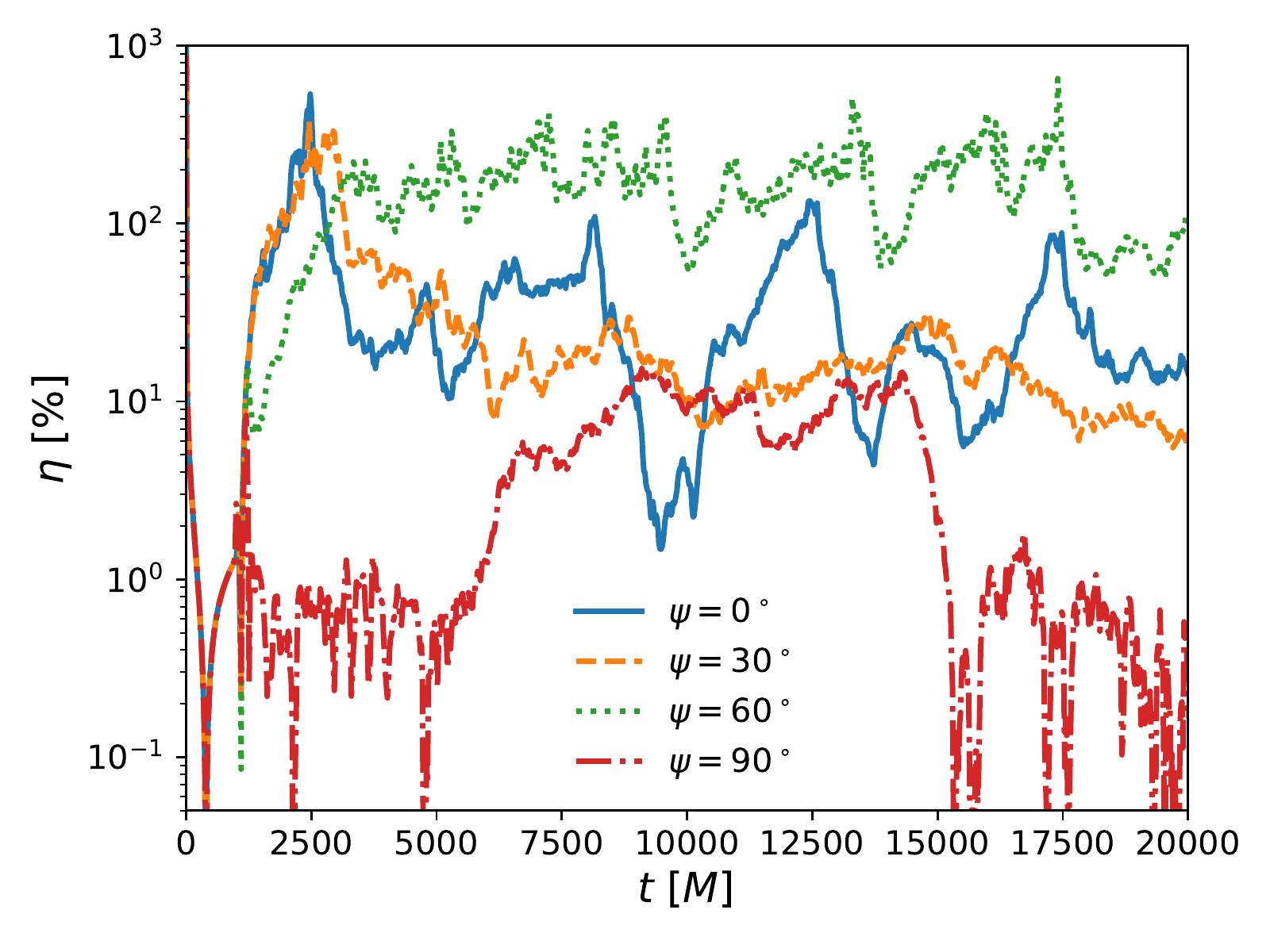}
\caption{Angle integrated quantities plotted vs. time for the four simulations.  Top: Dimensionless measure of the amount of magnetic flux threading the event horizon, $\phi_{\rm BH}$. Middle: Net mass accretion rate through the horizon normalized to the Bondi rate, $|\dot M/ \dot M_{\rm Bondi}|$.  Bottom: Outflow efficiency measured at $r = 5 r_{\rm g}$, $\eta \equiv |\dot E - \dot M|/|\dot M|$.  The $\psi = 0^\circ$ and $\psi = 60^\circ$ show magnetically arrested behavior, with flux dissipation events (i.e., rapid decreases in $\phi_{\rm BH}$) followed by large spikes in accretion rates.  The other two simulations are less time variable in both magnetic flux and accretion rate, saturating at moderately smaller values of $\phi_{\rm BH}$.  $\eta$ is dominated by the electromagnetic energy of the jet, and reaches as high as $200$--300\% (meaning that energy is extracted from the rotation of the black hole) for $\psi =60^\circ$.  The other three simulations have more modest efficiencies, $\sim$ 10--30\% on average.  This is true even of the $\psi=90^\circ$ simulation with no net initial vertical flux, though only for $5000M \lesssim 15000 M$ when a jet is present (see Figure \ref{fig:Bx_contour}). At other times $\eta$ is $<$ 1\%.  } 
\label{fig:mdot_time}
\end{figure}

We now turn our attention to the temporal variability of the four simulations.  A key quantity in magnetized black hole accretion is the dimensionless measure of the magnetic flux threading the horizon, 
\begin{equation}
  \phi_{\rm BH} \equiv \frac{\sqrt{4{\rm \pi}} \iint |B^r| d\Omega}{2 \sqrt{|\dot M|}},
\end{equation}
where $B^r$ is the radial component of the magnetic three-vector and $\dot M$ is the accretion rate. In GRMHD torus simulations this quantity seems to have a maximum possible value at $\phi_{\rm BH} \approx $ 40--60 (e.g., \citealt{Sasha2011,White2019}), at which point the flow transitions to the magnetically arrested state where the outwards Lorentz force is strong enough to periodically halt accretion in a highly time-variable configuration \citep{Igumenshchev2003,Narayan2003}.  In the top panel of Figure \ref{fig:mdot_time}, we plot $\phi_{\rm BH}$ for our four simulations.  The $\psi = 0^\circ$ curve displays quintessential magnetically arrested disk (MAD) characteristics, with $\phi_{\rm BH}$ varying between 30--60 in periodic cycles of accumulation and dissipation. The accretion rate in the second panel of Figure \ref{fig:mdot_time} varies accordingly (between 0.05--0.35 times the Bondi rate) with the peaks in $\phi_{\rm BH}$ associated with valleys in $\dot M$, and vice versa.   In contrast, the $\psi = 30^\circ$ simulation does not seem to reach the MAD state, with a saturated $\phi_{\rm BH}$ of $\sim $ 30 that is much less variable than the $\psi = 0^\circ$ case.  The accretion rate is also more constant in time than $\psi = 0^\circ$, though the time-averaged values in both simulations are comparable, $\sim$ 0.2 times the Bondi rate.  The $\psi = 90^\circ $ simulation has an even smaller amount of magnetic flux reaching the black hole, $\phi_{\rm BH} \approx 20$ and roughly constant in time. This makes sense since the jet seen in Figure \ref{fig:Bx_contour} pushes perpendicular to the initial magnetic field and effectively limits $\phi_{\rm BH}$.  Because of this, the time-averaged accretion rate is the largest of the four simulations, $\sim$ 0.25 times the Bondi rate.  The most a priori surprising case is $\psi = 60^\circ$, which shows the largest amount of magnetic flux reaching the black hole, $\phi_{\rm BH} \sim$ 60--100, as well as the smallest accretion rate, $\dot M $ $\sim$ 0.02--0.1 times the Bondi rate.  These curves display typical MAD behavior similar to the $\psi = 0^\circ$ curves.  Naively, we would have expected the results for $\psi = 60^\circ$ simulation to fall somewhere in between the $\psi = 30^\circ$ and the $\psi = 90^\circ$ results with a relatively small amount of magnetic flux reaching the black hole.  Instead, it seems that supplying a field that is significantly tilted from vertical ultimately results in more net flux (as measured in both the absolute sense and normalized to the square root of the accretion rate, $\phi_{\rm BH}$) than a field completely aligned with the spin of the black hole.  We speculate on why in \S \ref{sec:60}.

The third panel of Figure \ref{fig:mdot_time} plots the outflow efficiency, $\eta = (\dot E - \dot M) /|\dot M|$.   As expected, the simulation with the largest $\phi_{\rm BH}$ shows the largest efficiency ($\sim 200$ \% for $\psi = 60^\circ$), while the simulation with the smallest $\phi_{\rm BH}$ shows the smallest efficiency (either $\sim 1$ \% or $\sim$ 10 \% for $\psi=90^\circ$).  An efficiency of 200 \% can only occur if energy is being extracted from the spin of the black hole\footnote{For $\eta<100\%$ it is also possible that the flow is extracting energy from the black hole but it is not as easy to diagnose.}, something observed in previous highly magnetized GRMHD simulations (e.g., \citealt{Sasha2011}). The temporal variability of the efficiency in the $\psi = 90^\circ$ simulation differs from the other three simulations in that there are both quiescent phases and an active phase.  The active phase (with $\eta \sim 10$\%) occurs between $5000 M \lesssim t \lesssim 15000 M$, while the quiescent phases (with $\eta \sim 1$\%) occur for $t\lesssim5000 M$ and $t\gtrsim 15000 M$.  The quiescent phases correspond to the highly convective state discussed in \S \ref{sec:overview} where a lack of significant vertical flux prevents the formation of a jet, instead favoring disordered reconnection-driven turbulence (see top right and bottom left panel of Figure \ref{fig:Bx_contour}). The active phase arises after enough locally net vertical magnetic field is created via turbulence to form a jet.  

\subsection{Jet Properties}
\label{sec:jet}

\begin{figure*}
\includegraphics[width=0.95\textwidth]{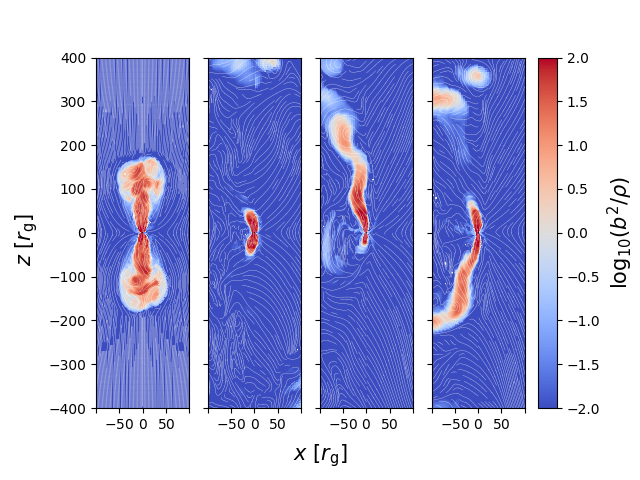}
\caption{Azimuthal slice of the magnetization parameter $b^2/\rho$ at four different times in our $\psi = 0^\circ$ simulation. Starting from the top left panel and proceeding to the right, the panels represent 2000, 8000, 14000, and 20000 $M$.   A highly magnetized, well-collimated jet is visible at all times, though its size and orientation are significantly time-variable.  It is also highly asymmetric about both the midplane and the polar axes, with the upper ($z>0$) and lower ($z<0$) portions of the jet often displaying very different behavior.   Each can be tilted by as much as $30^\circ$ from the black hole spin axis at certain times. An animated version of this figure is available at \href{https://youtu.be/c6tVZ5qyumA}{https://youtu.be/c6tVZ5qyumA}.} 
\label{fig:Bz_contour_jet}
\end{figure*}

\begin{figure*}
\includegraphics[width=0.95\textwidth]{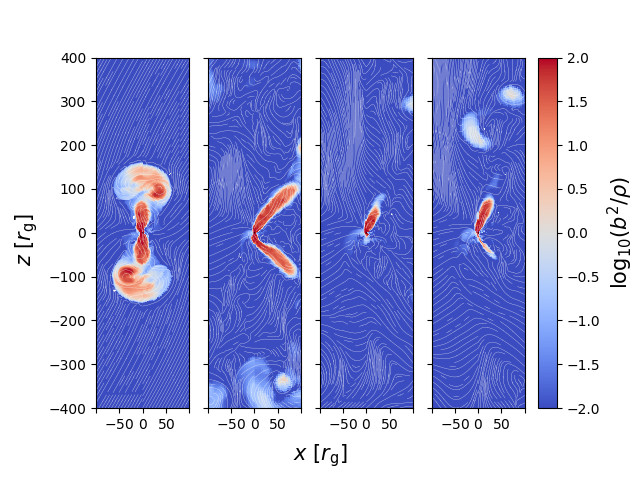}
\caption{Same as Figure \ref{fig:Bz_contour_jet} but for $\psi = 30^\circ$.} 
\label{fig:30_contour_jet}
\end{figure*}

\begin{figure*}
\includegraphics[width=0.95\textwidth]{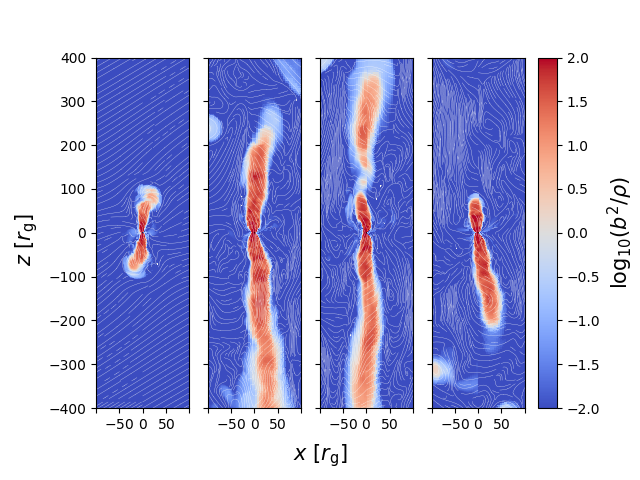}
\caption{Same as Figure \ref{fig:Bz_contour_jet} but for $\psi = 60^\circ$.} 
\label{fig:60_contour_jet}

\end{figure*}

\begin{figure*}
\includegraphics[width=0.95\textwidth]{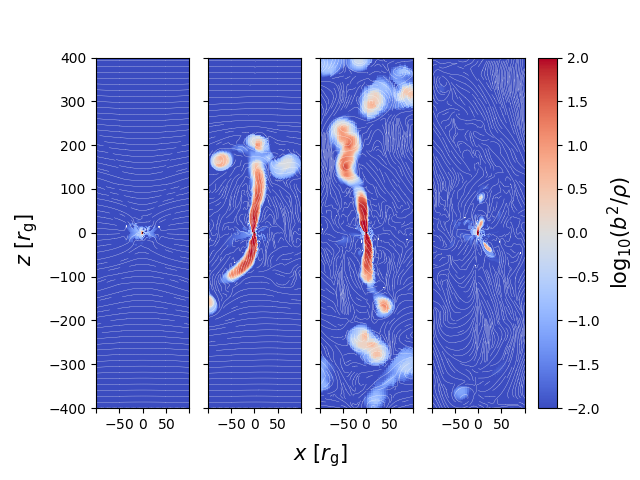}
\caption{Same as Figure \ref{fig:Bz_contour_jet} but for $\psi = 90^\circ$. } 
\label{fig:Bx_contour_jet}
\end{figure*}

In this section we focus specifically on the magnetically dominated jet.   Figures \ref{fig:Bz_contour_jet}--\ref{fig:Bx_contour_jet} show 2D slices at four different times of the magnetization parameter $b^2/\rho$ zoomed out to a $200 r_{\rm g}\times 800 r_{\rm g}$ box in the $x$--$z$ plane in the four simulations.  At early times (i.e., 2000 $M$ in the left panel of Figures \ref{fig:Bz_contour_jet}--\ref{fig:Bx_contour_jet}), the jets are relatively symmetric about both the spin axis of the black hole and the midplane.  As time goes on, however, this symmetry is broken by external kink instabilities that create more complicated structures (see, e.g. \citealt{Bromberg2016,Sasha2016,Rodolfo2017}).   At some times the jet is significantly more extended in either the upper or lower directions (e.g., the two rightmost panels of Figure \ref{fig:Bz_contour_jet} and Figure \ref{fig:30_contour_jet} as well as the rightmost panel of Figure \ref{fig:60_contour_jet}) and can favor the left or right side of the domain (e.g.,  the two rightmost panels fo Figure \ref{fig:Bz_contour_jet} and the three rightmost panels of Figure \ref{fig:30_contour_jet}). The latter phenomenon is particularly striking  in the second panel of Figure \ref{fig:30_contour_jet} ($\psi = 30^\circ$), where the upper and lower jet make an approximately 90$^\circ$ angle with each other.    While the upper jet is aligned with the direction of the initial magnetic field (inclined by $30^\circ$ from the spin axis), the lower jet is perpendicular to that direction and thus the alignment in the upper jet may just be a coincidence.  That conclusion is supported by the fact that the $\psi = 0^\circ$ jet also shows a non-neglible tilt away from the spin axis in the two rightmost panels of Figure \ref{fig:Bz_contour_jet} despite the field having been initially aligned with the spin axis.  It is likely that the degree to which the jet is tilted is determined by stochastic symmetry breaking as it propagates through a relatively dense medium and wobbles about via kink modes. 

The jets are also intermittent in the sense that there are spatial gaps between different sections (e.g., third panel of Figure \ref{fig:60_contour_jet} and Figure \ref{fig:Bx_contour_jet}) and their spatial extent periodically grows and diminishes.   The intermittency is caused by the periodic fluctuations of the fundamental power source of the jet, the magnetic flux, seen in the top panel of Figure \ref{fig:mdot_time}.  Without this power source, the cavities initially excised by the jet are quickly overcome by the surrounding gas as it falls inwards.

Another interesting aspect of Figures \ref{fig:Bz_contour_jet}--\ref{fig:Bx_contour_jet} are the bubbles that pinch off from the main body of the jet (e.g., the middle two panels of Figure \ref{fig:Bx_contour_jet} and rightmost panel of Figure \ref{fig:Bz_contour_jet}).  These are low density, magnetic pressure supported regions of tangled magnetic field that form as the jets propagate outwards through the ambient medium.  They buoyantly rise through the gas until they dissipate at larger radii.


\begin{figure}
\includegraphics[width=0.45\textwidth]{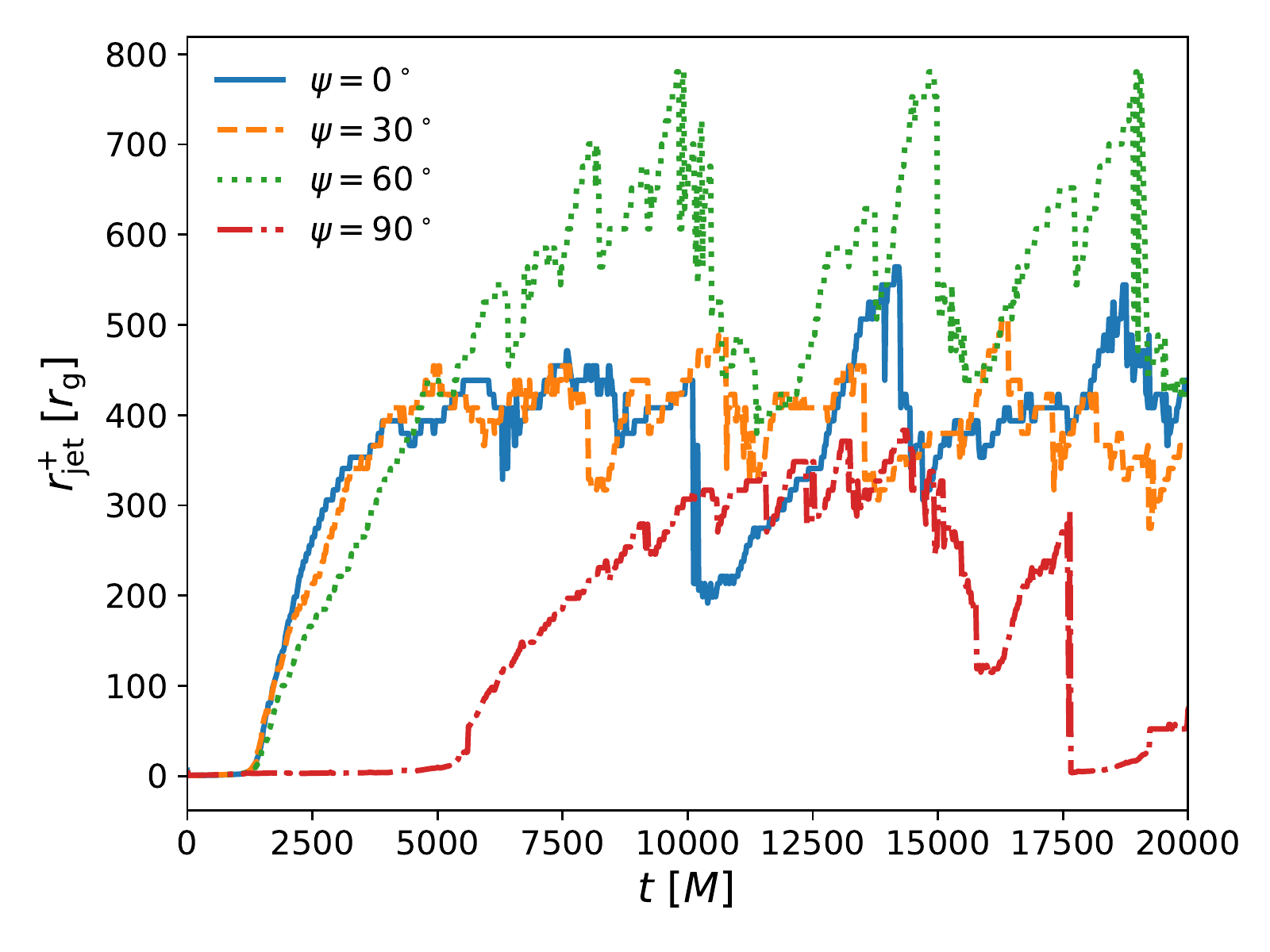}
\includegraphics[width=0.45\textwidth]{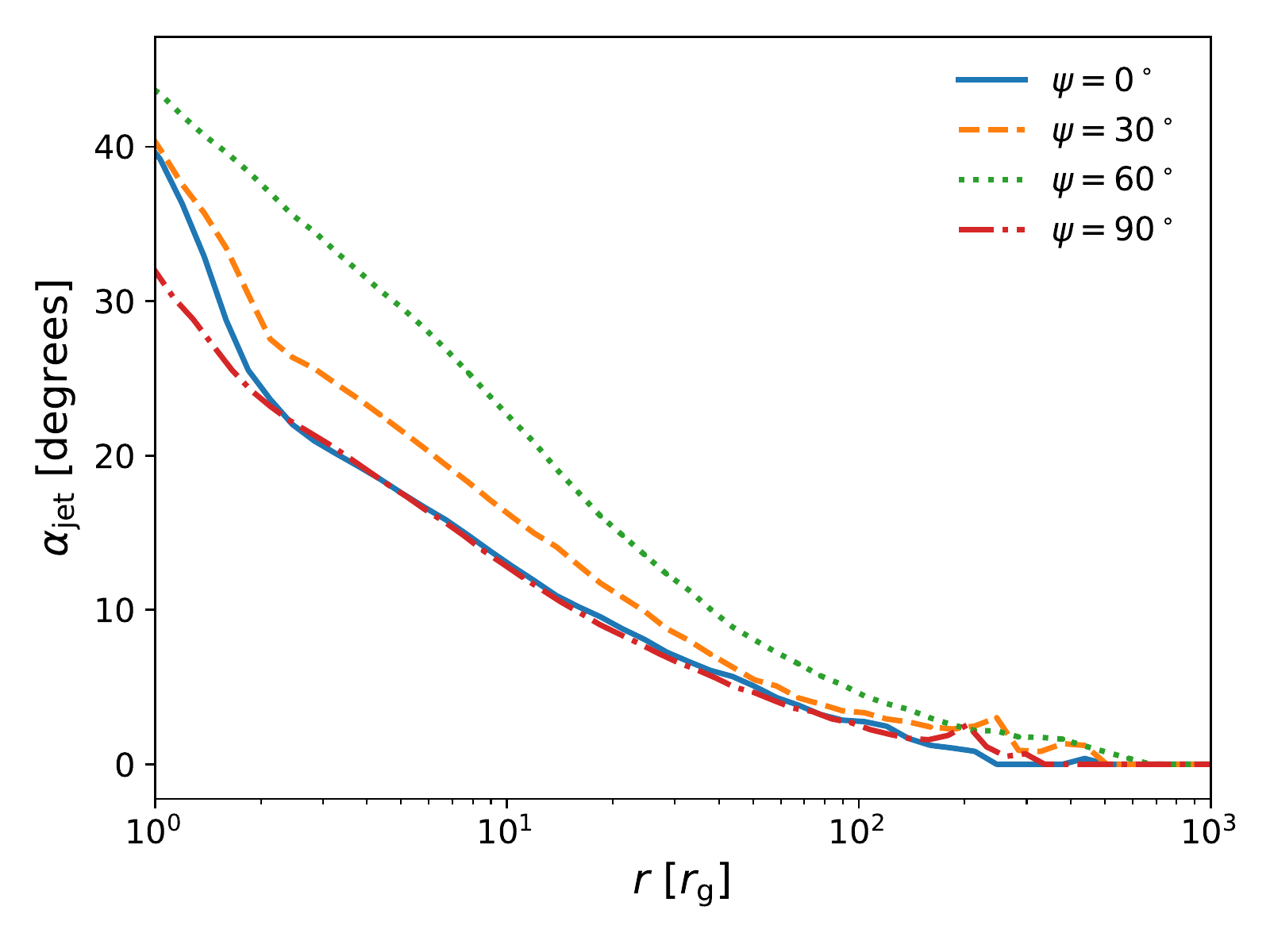}
\caption{Upper jet radius, $r^{+}_{\rm jet} $, as a function of time (top) and jet half-opening angle, $\alpha_{\rm jet}$, as a function of radius (bottom) in our four simulations.  Jets initially travel outwards at a constant speed that is a fraction of the speed of light but ultimately stall at $\sim$ 300--700$r_{\rm g}$, at which point they periodically fall back and then re-expand.   The opening angles of the jets decrease with radius in a manner consistent with a parabolic shape.  } 
\label{fig:jet_radius}
\end{figure}

In order to further quantify the properties of the jet, we define it precisely as all regions with $b^2/2 > 1.5 \rho $ (as in \citealt{Liska2019}).  This allows us to assign radii to the upper and lower jets, $ r_{\rm jet}^{+/-}$, as the maximum radii for which $b^2/2 > 1.5 \rho $ and $\theta<{\rm \pi}/2$ (+) or $\theta>{\rm \pi}/2$ (-).  Although these definitions are somewhat crude, they relatively accurately represent the behavior seen in Figures \ref{fig:Bz_contour_jet}--\ref{fig:Bx_contour_jet} so they suit our purposes.  $r_{\rm jet}^{+}$ is plotted vs. time for the four simulations in the top panel Figure \ref{fig:jet_radius} (the qualitative behavior of $r_{\rm jet}^{-}$ is similar).  Initially, the jets expand outwards at a fraction of the speed of light but eventually stall around $300$--$800$ $r_{\rm g}$. 
At these radii the surrounding gas is roughly uniform and stationary since the Bondi radius is $200 r_{\rm g}$.  After stalling, the jet radius oscillates in time, correlated directly with fluctuations in $\phi_{\rm BH}$ and $\eta$ in Figure \ref{fig:mdot_time}.   The differences between the time averaged jet radius in the four simulations are as expected based on the differences in outflow efficiency  (Figure \ref{fig:mdot_time}).  The $\psi = 90^\circ$ jet stalls at the smallest radius, $\sim$ $300 r_{\rm g}$, while the $\psi = 60^\circ$ jet stalls at the largest radius, $\sim$ 700--800$r_{\rm g}$.  The $\psi = 0^\circ$ and $\psi = 30^\circ$ jets stall somewhere in between, $\sim $ $ 400 r_{\rm g}$.

We find that velocities inside the jet are only mildly relativistic, reaching maximum Lorentz factors $\gamma = u^t ( - g^{tt})^{-1/2} \lesssim 2$ ($u^r/u^t \lesssim 0.6 c$) at a few hundred $r_{\rm g}$ and then rapidly decelerating in radius near the edge of the jet.  

We can also define a half opening angle of the jet, $\alpha_{\rm jet}$ as the solution to the equation
 \begin{equation}
\frac{A_{\rm jet}(r)}{2{\rm \pi}} - \int\limits_0^{\alpha_{\rm jet}} \gdet d \theta =  0,
\end{equation}
where $A_{\rm jet}(r) \equiv 1/2 \iint H(b^2/2 - 1.5 \rho) \gdet d\theta d\varphi $  is the effective cross-sectional area of one jet (hence the factor of 1/2) and $H$ is the Heaviside step function.   The time-averaged $\alpha_{\rm jet}$ is plotted vs. radius in the bottom panel of Figure \ref{fig:jet_radius} for our simulations.  Near the event horizon, the jets are quite wide, $\alpha_{\rm jet}$ $\sim$ 30--40$^\circ$ as shown previously in Figures \ref{fig:Bz_contour}--\ref{fig:Bx_contour}.  As material in the jet flows outwards it becomes confined to a narrower and narrower angle, decreasing in a way that is consistent with a parabolic jet profile ($A_{\rm jet} $ $\tilde \propto $ $r $). For example, at $r=10 r_{\rm g}$ $\alpha_{\rm jet}$ $\sim$ 10--20$^\circ$ while at $r = 100 r_{\rm g}$ $\alpha_{\rm jet}$ $\sim$ a few degrees.  

\begin{figure}
\includegraphics[width=0.45\textwidth]{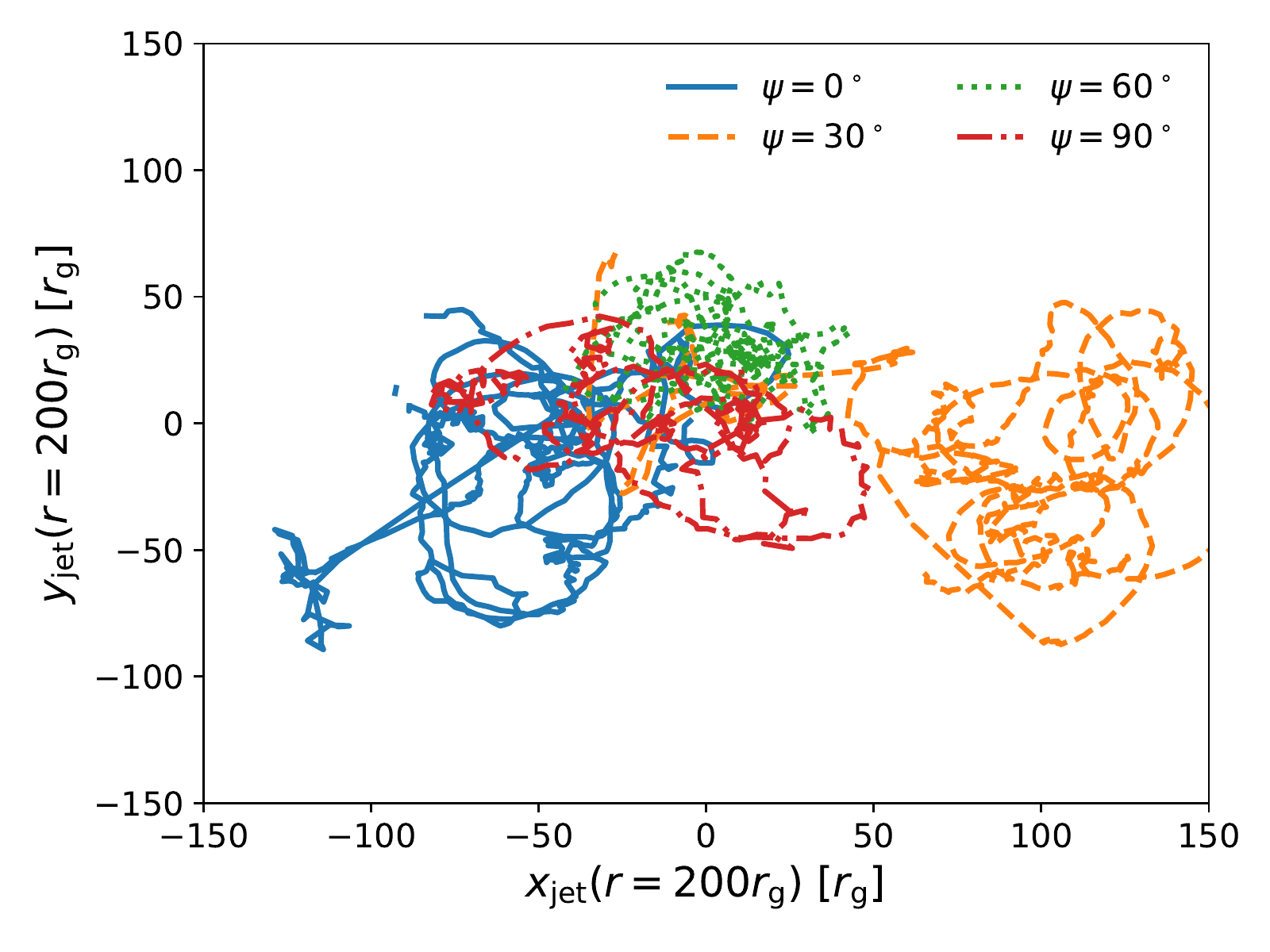}
\caption{Space-time diagrams of the location of the jet at $r = 200 r_{\rm g}$ projected onto the $x-y$ plane in the four simulations.  The kink instability causes the jets to move around in spiral-like patterns as a function of time.   With the largest jet power (Figure \ref{fig:mdot_time}), the $\psi = 60^\circ$ simulation also has smallest amplitude variations in jet location.} 
\label{fig:jet_spacetime}
\end{figure}

\begin{figure}
\includegraphics[width=0.45\textwidth]{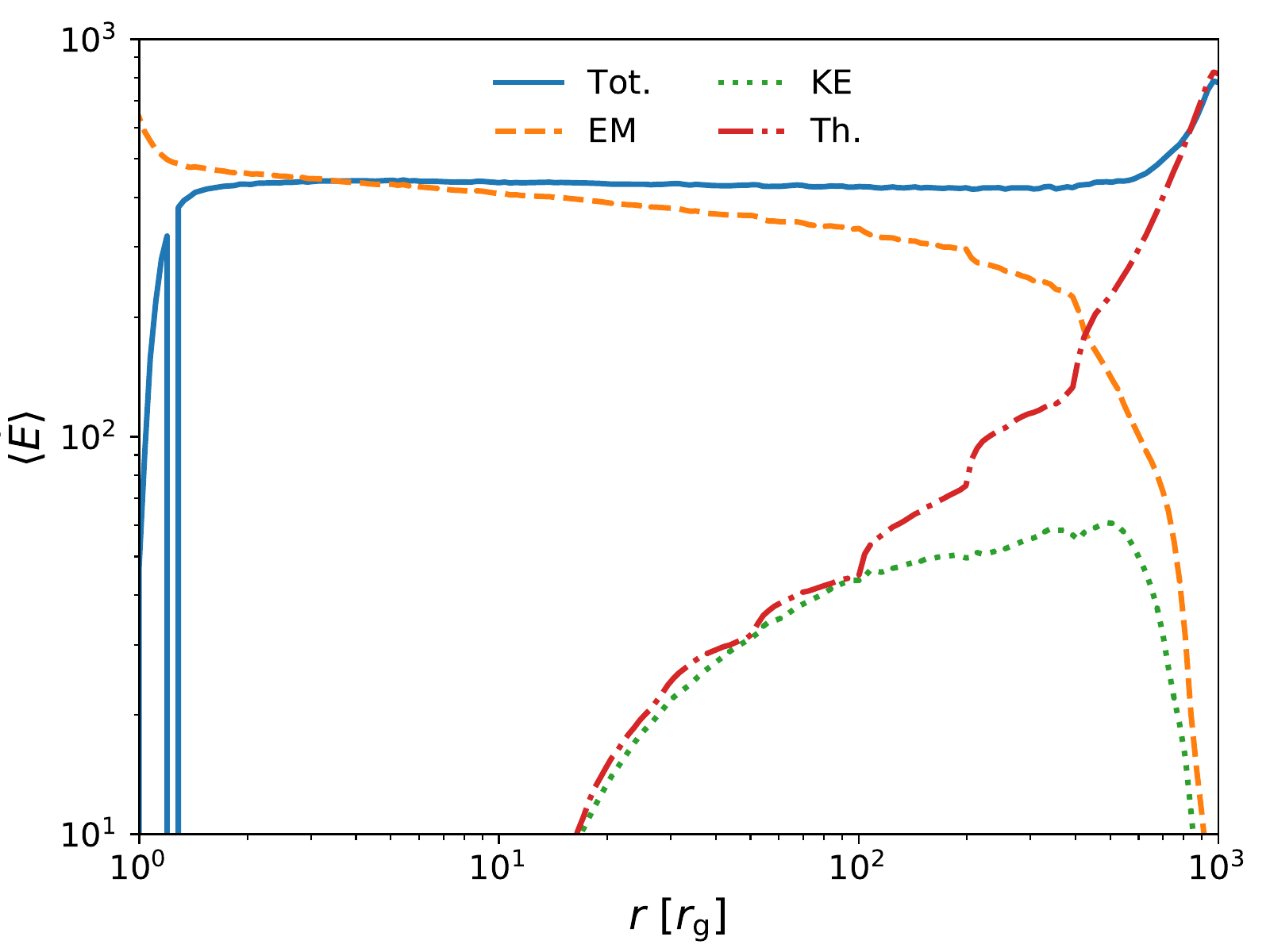}
\caption{Radial energy flux, $\dot E$, broken up into total (solid), electromagnetic (dashed), kinetic (dotted), and thermal (dot-dashed) components for the $\psi = 60^\circ$ simulation.  At small radii, the energy outflow is dominated by the Poynting dominated jet.  As the jet propagates outwards, its magnetic energy dissipates into kinetic and thermal energy.  The latter dominates for $r\gtrsim 500 r_{\rm g}$.  The other three simulations show qualitatively similar behavior and are thus not shown.} 
\label{fig:Edot_jet}
\end{figure}

\begin{figure*}
  \begin{center}
\includegraphics[width=0.95\textwidth]{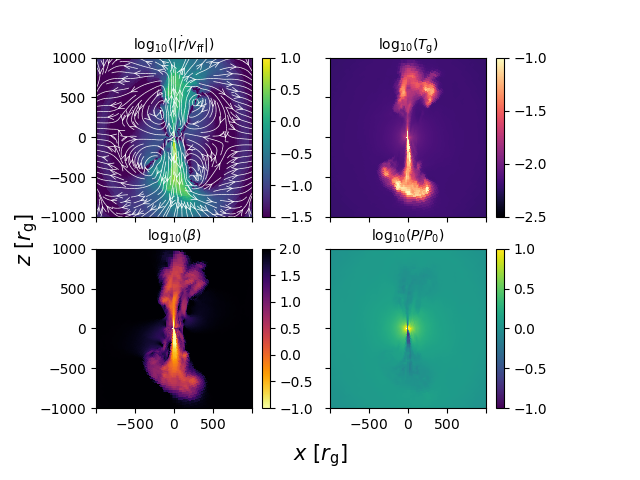}
\end{center}
\caption{Time-averaged, $y=0$ slices of four quantities in the $\psi = 60^\circ$ simulation at large radii, where the jet breaks up due to the kink instability.  Top left: magnitude of the radial velocity normalized to the free-fall speed, $|\dot r|/v_{\rm ff}$, overplotted with velocity streamlines.  Top right: gas temperature, $T_{\rm g}$.  Bottom left: plasma $\beta$.  Bottom right: pressure normalized to the initial pressure, $P/P_0$.  The ``wobbling'' of the jet (Figure \ref{fig:jet_spacetime}) caused by the kink instability carves out a cavity of hot, magnetized gas with large radial velocities.  This cavity is in rough pressure equilibrium with its surroundings.    Gas near the edge of the jet/cavity circulates back into the inflow, altering the supply of magnetic flux in a nonlinear way.  The other three simulations show qualitatively similar behavior and are thus not shown.} 
\label{fig:jet_velocity}
\end{figure*}

Figure \ref{fig:jet_spacetime} shows spacetime diagrams of the location of the jets (at $r=200 r_{\rm g}$) projected onto the $x$--$y$ plane.   More precisely, we define 
\begin{equation}
x_{\rm jet}^{+} = \frac{\iint x H(b^2/2 - 1.5 \rho )H(\theta<{\rm \pi}/2)   \gdet d\theta d\varphi}{\iint H(b^2/2 - 1.5 \rho ) H(\theta<{\rm \pi}/2)   \gdet d\theta d\varphi}
\end{equation}
and similarly for $y_{\rm jet}^{+}$.  The jets wobble around with time in spiral patterns characteristic of the external kink instability, which is known to occur when a highly magnetized jet attempts to penetrate a dense medium (e.g., \citealt{Bromberg2016,Sasha2016,Rodolfo2017}) and is generally caused by toroidally dominant fields in collimated jets \citep{Begelman1998,Lyubarskii1999}.   As the jets in our simulations push against the surrounding gas, they decelerate, converting some of their poloidal magnetic field to toroidal field like a spring being compressed, leading to the unstable configuration near the radial edge of the jet.   As a result, its electromagnetic energy is largely converted into internal and kinetic energy by heating up and accelerating the surrounding gas.  Figure \ref{fig:Edot_jet} demonstrates this for the $\psi = 60^\circ$ simulation by plotting the time averaged $\dot E$ vs. radius, broken up into electromagnetic, kinetic, and thermal components.  The thermal component starts off as negligible but steadily increases until it overtakes the electromagnetic component at $\approx $ $500 r_{\rm g}$ (typically $\sim $ a few 100 $r_{\rm g}$ for the other three simulations with smaller jet power).   At this point the jet has blown out a low density, high temperature cavity of much larger spatial extent than its cross-section.  This is shown in Figure \ref{fig:jet_velocity}, which plots the time averaged radial velocity, temperature, $\beta$, and pressure in the $x$--$z$ plane for the same $\psi = 60^\circ $ simulation on an $r \sim $ 1000 $r_{\rm g}$ scale.   Compared to the surrounding medium, the cavity is a factor of $\sim$ 10 times hotter and $\sim$ 10 times less dense so that the pressures are comparable.  Most of the gas in the cavity is unbound, with velocities exceeding free-fall.

Also plotted in the top left panel of Figure \ref{fig:jet_velocity} are velocity streamlines.  Material on the edge of the jet/cavity with $\dot r$ $\lesssim$ free-fall
tends to circulate back towards the midplane and feed the inflow.  In practice, while this does not significantly alter the hydrodynamic properties of the accretion flow near the Bondi radius, it can noticeably change the field geometry there (see, e.g., the rightmost panel of Figure \ref{fig:60_contour_jet} where the field lines in the $\psi = 60^\circ$ simulation bear almost no resemblance to the initial conditions and look more like the $\psi = 0^\circ$ simulation).    Because of this effect we are less concerned that the idealized, purely laminar magnetic fields lines in the initial conditions are artificially reducing turbulence in the midplane.



\section{The $\psi = 60^\circ$ Outlier}
\label{sec:60}

\begin{figure}
\includegraphics[width=0.45\textwidth]{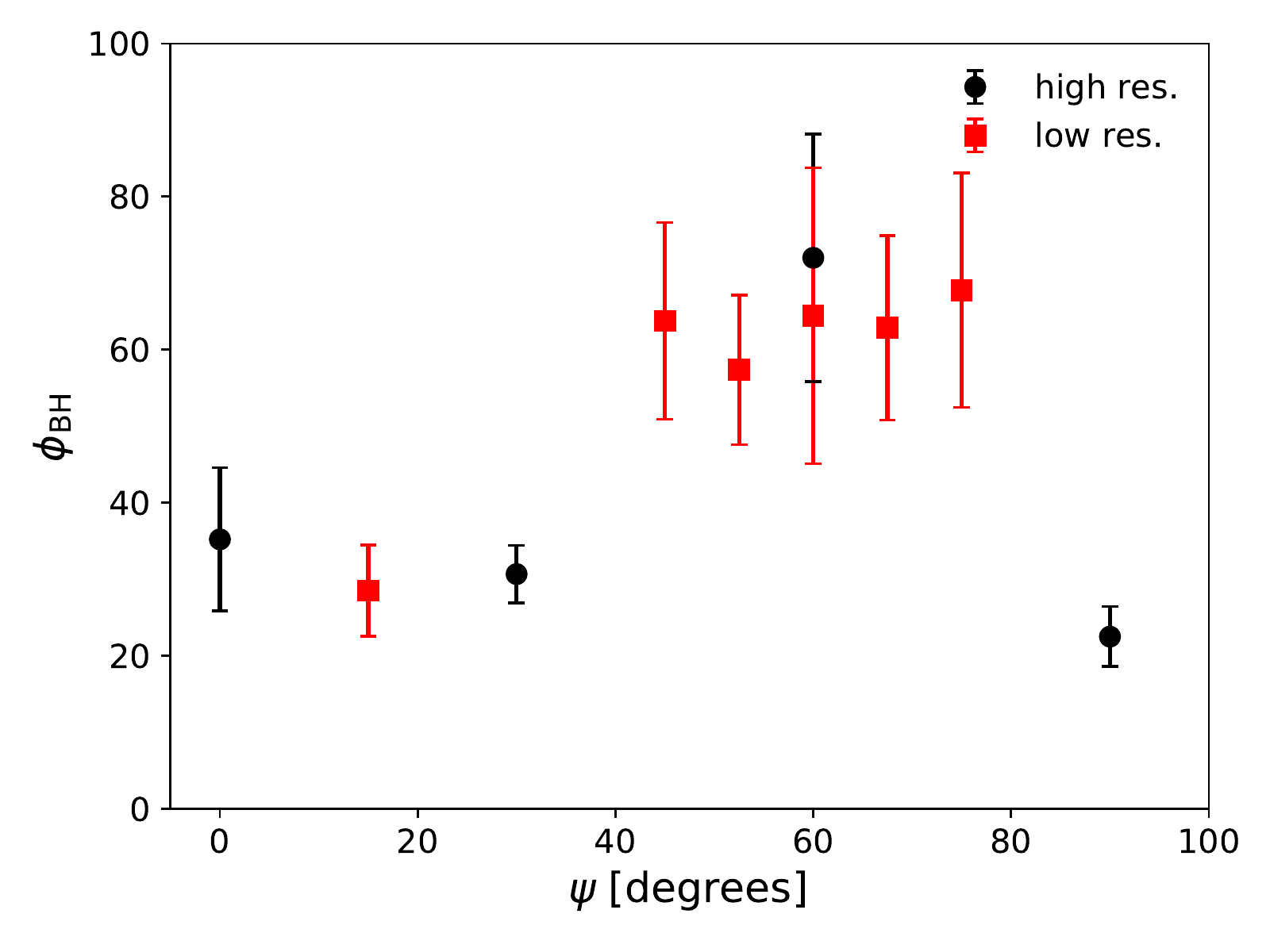}
\caption{Dimensionless magnetic flux threading the black hole averaged over time (see main text), $\phi_{\rm BH}$, vs. initial magnetic tilt angle $\psi$. Circles represent simulations that employ the fiducial resolution described in \S \ref{sec:model}, squares represent simulations that employ a resolution reduced by 2 in all three directions, while error-bars represent the standard deviation of $\phi_{\rm BH}$ over the chosen time interval.  There is a clear peak at $\psi \sim 40^\circ$--$80^\circ$ that we discuss in \S \ref{sec:60}.  
} 
\label{fig:phi_vs_psi}
\end{figure}

Naively, one would assume that the amount of magnetic flux threading the black hole (and thus jet power) would either decrease with increasing $\psi $ (for $0^\circ \le \psi  \le 90^\circ $) because of the decreased supply of initial vertical magnetic field or be relatively independent of $\psi$ because $\phi_{\rm BH}$ would tend to saturate at some unique value determined by the balance between the outwards Lorentz force and inflow.   Instead, we find that the $\psi =60^\circ$ simulation has the highest value of $\phi_{\rm BH}$.   To determine whether this result is robust or simply caused by stochastic variation between the four simulations, we run an additional six simulations with $\psi = 15^\circ, 45^\circ,52.5^\circ,60^\circ,67.5^\circ$, and $75^\circ$ with exactly half the resolution in each direction as the main simulations of this paper.  The resulting $\phi_{\rm BH}$ as function of $\psi$ is plotted in Figure \ref{fig:phi_vs_psi} (including all 10 simulations), where $\phi_{\rm BH}$ is averaged over intervals containing several limit cycles once the flux has roughly saturated.\footnote{For most simulations we use the interval 5000 $M$ --20000 $M$, but for some of the lower resolution simulations the flux takes longer to saturate and so we use intervals with later start times.  For example, for $\psi=45^\circ$ we use 16000 $M$ --25000 $M$}  With the range in $\psi$ now finely sampled, it is clear that simulations with $40^\circ \lesssim \psi \lesssim80^\circ$ have systematically higher values of $\phi_{\rm BH}$ than the rest ($\sim$ 60--80 compared to $\sim$ 20--40).  

\begin{figure*}
\includegraphics[width=0.95\textwidth]{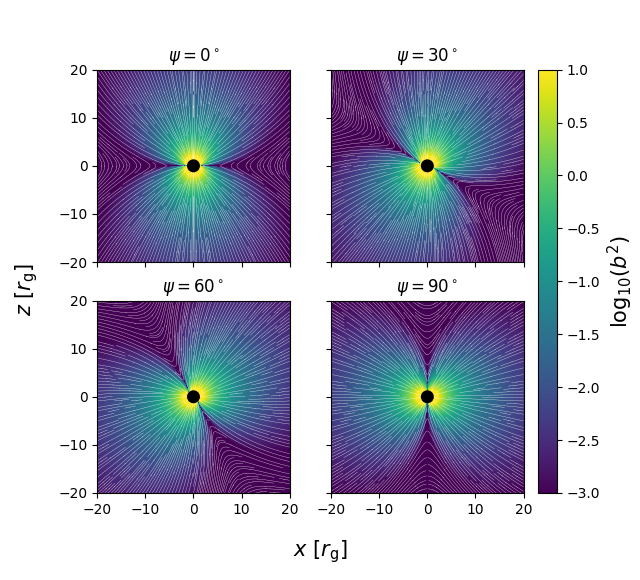}
\caption{2D slice in the $x$-$z$ plane of $b^2$ overplotted with magnetic field lines at an early time ($t = 400 M$) in our four simulations.   Spherical advection of the magnetic field causes a current sheet of $b^2 \approx 0$ to form perpendicular to the initial magnetic field. On one side of the current sheet field lines are mostly radial and pointed outwards, while on the other side field lines are mostly radial and pointed inwards.  The current sheet itself is the location of the first reconnection events that drives turbulence.  For larger tilts (i.e., $\psi = 60^\circ$ and $\psi = 90^\circ$), it overlaps with the polar regions that are ultimately evacuated by jets.  This means that if the jet forms before the first reconnection event (as it does for $\psi = 60^\circ$ but not $\psi = 90^\circ$), there can be a reduced amount of turbulence and an increase in the amount of magnetic flux that reaches the event horizon.       } 
\label{fig:bsq_contour}
\end{figure*}

To understand the dependence of $\phi_{\rm BH}$ on $\psi$ shown in Figure \ref{fig:phi_vs_psi}, first consider the case of a non-spinning black hole.  In the limit of large Bondi radius, the initially uniform, weak magnetic field become almost entirely radial by the time it reaches the event horizon, flipping sign across a current sheet in the plane perpendicular to the initial field and containing the origin, as illustrated in Figure \ref{fig:bsq_contour} which plots $b^2$ and magnetic field lines at an early time in our simulations (at this early time the effects of $a \ne 0$ are small). The field maintains this geometry all the while growing in strength until it becomes dynamically important \citep{Bisnovatyi1974}.  At this point, it reconnects at the location of the current sheet, which heats the gas and drives turbulence.  Turbulence leads to more reconnection, and the cycle continues.   Rotating the initial magnetic field simply rotates the location of the initial current sheet so that the resulting flows are essentially identical, modulo the chaotic nature of the turbulence.    Now consider a rapidly spinning black hole.  Once a jet is formed the inflow becomes restricted to a smaller region centered on the midplane (as defined by the black hole rotation axis), say, $|\theta-{\rm \pi/2}|<\alpha_{\rm mid}$ that will only contain the entire angular extent of the initial current sheet if $\psi <({\rm \pi}/2-\alpha_{\rm mid})$ (assuming $\psi <90^\circ$).   If the jet forms before the onset of turbulence then the inflow in that case would plausibly be characterized by more reconnection (and hence, more turbulence and more dissipation of the magnetic field) than the case when the initial current sheet is only partially contained in the inflow region (the rest being blown away by the jet).  Less dissipation in the more laminar flows would allow for larger amount of magnetic flux to accumulate and a higher saturation value for $\phi_{\rm BH}$.  This argument breaks down if the field reconnects and starts driving turbulence \emph{before} the jet has a chance to form, which happens when $\psi $ nearly approaches $90^\circ$ since it takes longer for vertical field to accumulate via advection.  Thus there is a `sweet spot' around ${\rm \pi}/2 - \alpha_{\rm mid} <\psi < 90^\circ$ where we would expect $\phi_{\rm BH }$ to be largest.  Since we find $({\rm \pi}/2 - \alpha_{\rm mid}) \gtrsim 45^\circ$ near the horizon (bottom panel of Figure \ref{fig:jet_radius}), this would explain why the $\psi = 60^\circ$ simulation has the highest value of $\phi_{\rm BH}$.   


\section{Dependence of Our Results on Certain Parameters}
\label{sec:param}
In this section we speculate how the results of our simulations depend on certain input parameters, namely, the Bondi radius, the initial $\beta$, and the maximum magnetization, $\sigma_{\rm max}$.  Although it is beyond the scope of this work to do a comprehensive parameter survey, we can make reasonable extrapolations based on physical arguments and the results of past work in the literature.  Even so, this is no substitute for systematically varying these parameters in our simulations and thus the conclusions we draw in this section should be treated as conjecture.

\subsection{Dependence on Bondi Radius}
Our choice of $r_{\rm B} = 200 r_{\rm g}$ is much less than appropriate for realistic low luminosity AGN such as Sgr A* and M87.  In practice this limits the region of reconnection-driven turbulence in our simulations to a relatively small radial range.   Since an initially uniform, weak magnetic field will grow in strength due to the compression caused by spherical inflow as $(r/r_{\rm B})^{-2}$ (dominated by the radial component), the magnetic pressure will grow as $(r/r_{\rm B})^{-4}$ while the thermal pressure will grow as $\sim$ $r^{-5/2}$ (assuming a Bondi type flow).  This means that  $\beta$ $\sim$ 1 will be reached at some radius $r_{\rm conv} \approx r_{\rm B} \beta_0^{2/3}$ where the field will start reconnecting.  This expression gives $r_{\rm conv} \approx 22 r_{\rm g}$ for our parameters, roughly agreeing with the simulations (Figure \ref{fig:radial_plots}).  For much larger $r_{\rm B}$, $r_{conv}$ will also be much larger and so we expect that convection will be present across several orders of magnitude in radius.   In this subsection, for the purposes of extrapolation we assume that the basic power law dependences of the flow properties that we observe in these simulations where convection/turbulence occurs in a relatively narrow radial range will hold for simulations where convection/turbulence occurs in a more extended radial range.  This assumption requires that a significant amount of net magnetic flux will survive transport through the larger convective regions and still reach the black hole.   Because the competition between advection and diffusion of magnetic fields in such a regime is not well understood, it is not obvious how well this assumption is justified.  The crux of the issue is that turbulent diffusion could, in principle, decouple the motion of the magnetic flux from the mass flux so that infall does not necessarily guarantee an accumulation of $\phi_{\rm BH}$ onto the black hole.


\subsubsection{$\phi_{\rm BH}$ vs. $\psi$ Dependence}
Also uncertain is whether or not the $\phi_{\rm BH}$ vs. $\psi$ dependence shown in Figure \ref{fig:phi_vs_psi} will hold for larger convective regions.  Our hypothesis for why the $\psi =60^\circ$ simulation has the largest $\phi_{\rm BH}$ (\S \ref{sec:60}) is that a large fraction of the initial current sheet is close enough to the polar regions to be partially blown away by the jet.   If the onset of convection/turbulence begins at much larger radii this subtlety of the magnetic field geometry may be washed out by the time it reaches the event horizon even if the overall net direction is preserved.  Another possibility is that the jet could stall at smaller radii than $r_{\rm conv}$ and have less of an influence on flux transport/reconnection. These concerns prevent us from determining whether the dependence of $\phi_{\rm BH}$ on $\psi$ seen in our models is a robust feature of spherical infall of initially coherent magnetic fields or unique to relatively small $r_{\rm B}$.   Future work varying $r_{\rm B}$ could shed more light on this issue.


\label{sec:bondi_ext}
\subsubsection{Accretion Rate}
Since we expect that the radial velocity near the event horizon to always be $\propto$ free fall $\sim$ $c$, the $\rho $ $\tilde \propto$ $ \rho_0 (r_{\rm B}/r)$ observed in our simulations implies $\dot M \propto\rho_0 r_{\rm B}$.  Since $\dot M_{\rm Bondi} \propto \rho_0 r_{\rm B}^{3/2}$, we predict that $\dot M/\dot  M_{\rm Bondi} \propto r_{\rm B}^{-1/2}$.  
For Sgr A*, where $r_{\rm B} \approx 2 \times 10^{5} r_{\rm g}$ and $\dot M_{\rm Bondi} \approx 10^{-6} $ $\dot M_{\odot}$/yr \citep{Baganoff2003,Wang2013}, this extrapolation would imply that $\dot M \sim$ $1.5 \times 10^{-3}$--$6 \times 10^{-3}$  $\dot  M_{\rm Bondi} $ $\sim$ 1.5--6 $\times 10^{-9}$ $\dot M_{\odot}$/yr, a reasonable accretion rate given observational constraints \citep{Marrone2006} and previous simulation-based estimates \citep{Shcherbakov2010,Mosci2014,Ressler2020b,Dexter2020}. 
For M87, where $r_{\rm B} \approx 5 \times 10^{5} r_{\rm g}$ and $\dot  M_{\rm Bondi} $ $\approx$ 0.1 $\dot M_{\odot}$/yr \citep{Dimatteo2003}, we extrapolate that $\dot M \sim$ $1 \times 10^{-3}$--$4 \times 10^{-3}$  $\dot  M_{\rm Bondi} $ $\sim$ 1--4 $\times 10^{-4}$ $\dot M_{\odot}$/yr, a number in good agreement with the mean accretion rates required for GRMHD MAD models to reproduce the 230 GHz flux \citet{EHT5} and consistent with Faraday rotation estimates (\citealt{Kuo2014}, though see \citealt{Ricarte2020} for a discussion of why the rotation measure might not be a reliable tracer for the accretion rate).  

Physically, the reason that this extrapolation predicts much smaller accretion rates than the Bondi value is not primarily due to mass outflows (though these are present near the jet boundary) but the reduced radial velocity caused by the presence of magnetic fields.  The Bondi rate is appropriate for a spherically symmetric distribution of gas under the influence of gravity alone and can thus be thought of as an effective upper limit on the accretion rate for a system that includes other forces (e.g., magnetic forces), irreversible dissipation (e.g., shocks or turbulence), or non-ideal effects (e.g., conduction).  

\subsubsection{Jet radius}

The radial extent of the jet is regulated by the kink instability, which causes the outer parts of the jet to spiral around the axis (Figure \ref{fig:jet_spacetime}) and generally become disrupted (Figures \ref{fig:Bz_contour_jet}--\ref{fig:Bx_contour_jet}).  This motion serves to dissipate the magnetic energy of the jet, heating the surrounding gas and propelling it into a wider, uncollimated outflow.  The transition happens when the timescale for kink mods to grow, $t_{\rm kink}$, is sufficiently small compared to the time it takes for a fluid element to traverse the remaining height of the jet, $t_{\rm dyn}$.  Using an analytic model of a cylindrical, relativistic jet propagating through an external medium motivated by their simulations, \citeauthor{Bromberg2016} (\citeyear{Bromberg2016}, see also \citealt{Bromberg2011}) estimate that 
\begin{equation}
  \label{eq:kink}
\frac{t_{\rm kink}}{t_{\rm dyn}} \tilde \propto \left(\frac{L_{\rm jet}}{\rho_{\rm a} r^2 \gamma_{\rm jet}^2}\right)^{1/6},
\end{equation}
where $L_{\rm jet}$ is the luminosity of the jet, $\rho_{\rm a}(r)$ is the density of the ambient medium, and $\gamma_{\rm jet}$ is the Lorentz factor of the jet.  
First consider $r< r_{\rm B}$, where $\rho_{\rm a}$ $\tilde \propto$ $r_{\rm B}/r$.  
Assuming $L_{\rm jet} \propto \dot E \propto \dot M c^2\propto r_{\rm B}$, Equation \eqref{eq:kink} predicts that the stability properties of the jet will be independent of $r_{\rm B}$ in in this region (as long as $\gamma_{\rm jet}$ is roughly independent of $r_{\rm B}$) and $t_{\rm kink}/t_{\rm dyn} \propto (r/r_{\rm g})^{-1/6}$.  
On the other hand, for $r>r_{\rm B}$, where the jets in our simulations stall, $\rho_{\rm a} \approx \rho_0$ and thus $t_{\rm kink}/t_{\rm dyn} \propto [r^2/(r_{\rm B} r_{\rm g})]^{-1/6}$.  

In order to estimate where a jet would ultimately stall if $r_{\rm B}$ was $\gg 200 r_{\rm g}$, we can thus solve 
\begin{equation}
 \begin{aligned}
   \left. \frac{t_{\rm kink}}{t_{\rm dyn}} \right|_{\rho_{\rm a} = \rho_0(r_{\rm B}/r)} \left(r=r_{\rm jet} ,r_{\rm B} \gg 200 r_{\rm g} \right)  \\
   =\left. \frac{t_{\rm kink}}{t_{\rm dyn}} \right|_{\rho_{\rm a} = \rho_0} \left(r= 400 r_{\rm g} ,r_{\rm B} = 200 r_{\rm g}\right) 
   \label{eq:rjet_est}
   \end{aligned}
\end{equation}
for $r_{\rm jet}$, where we have used $400 r_{\rm g}$ as a characteristic jet stalling radius for our simulations (Figure \ref{fig:jet_radius}) and approximated the density profile as a piecewise power law. Substituting Equation \eqref{eq:kink} for $t_{\rm kink}/t_{\rm dyn}$ into Equation \eqref{eq:rjet_est} we find $r_{\rm jet} \approx 800 r_{\rm g}$ independent of $r_{\rm B}$, which is valid as long as $r_{\rm B} \gtrsim 800 r_{\rm g}$.  For $r_{\rm B} \lesssim 800 r_{\rm g}$, we replace $\rho_{\rm a} =\rho_0 (r_{\rm B}/r)$ with $\rho_{\rm a} =\rho_0 $ in the right hand side of Equation \eqref{eq:rjet_est} and obtain $r_{\rm jet} \approx 400 r_{\rm g} \sqrt{r_{\rm B}/200 r_{\rm g}} $ .  In summary, we (roughly) predict
\begin{equation}
  r_{\rm jet} \approx \begin{cases} 
      800 r_{\rm g} \sqrt{\frac{r_{\rm B} }{800 r_{\rm g} } }  & r_{\rm B} \lesssim 800 r_{\rm g} \\
       800 r_{\rm g} & r_{\rm B} \gtrsim 800 r_{\rm g}.
   \end{cases}
   \label{eq:rjet_sol}
\end{equation}

In Sgr A* and M87, $r_{\rm B} \gg 800 r_{\rm g}$, so we expect that the jet would stall somewhere around $\sim$ 800 $r_{\rm g}$, $\pm$ a few hundred $r_{\rm g}$. This assumes, of course, that the linear stability criterion (Equation \ref{eq:kink}) correctly predicts the scaling of the non-linear disruption of the jet by the kink instability as $r_{\rm B}$ is varied.   


\subsection{Dependence on Initial $\beta$}
\label{sec:beta}
All simulations that we have presented use an initial $\beta_0=100$.  We do not expect our results to be sensitive to this value as long as it is sufficiently large ($\gtrsim 10$ or so) so that the initial magnetic field is not dynamically important.  This is because no matter how small the field is at $t=0$, the roughly spherical accretion of frozen-in field lines will lead to a configuration of nearly radial magnetic field lines near the horizon with a dependence on radius of $b^2 \propto r^{-4}$ and a magnitude that is growing with time \citep{Bisnovatyi1974}. The field strength will continue to grow until it becomes dynamically important, at which point the reconnection of oppositely directed field lines heats the gas and drives turbulence (e.g., \citealt{Shvartsman1971,Meszaros1975}).  In non-relativistic simulations,  \citet{Pang2011} found that their results were relatively insensitive to $\beta_0$ over the range 10--1000, with the density power law index $p$ having only a weak dependence on this parameter, $\beta_0^{-0.098}$.

\subsection{Dependence on $\sigma_{\rm max}$}
\label{sec:sigma}
As described  in \S \ref{sec:model}, we use a density floor to impose $\sigma\le \sigma_{\rm max} = 100$.  While this limit presumably has little to no effect on the main body of the accretion flow (where $\sigma$ is generally $\ll 1$), it effectively sets the value of $\sigma$ at the base of the jets in our simulations.  In reality $\sigma$ in these regions would be larger, though its precise value is unknown and may be set by pair-production processes \citep{BZ1977,Chen2018}.  A larger $\sigma$ in the base of the jet (i.e., a lower mass density) would lead to a larger Lorentz factor throughout the main body of the jet, which could affect the kink stability criterion and, consequently, $r_{\rm jet}$.   Equation \eqref{eq:kink} predicts, however, that the dependence of $t_{\rm kink}/t_{\rm dyn}$ on $\gamma_{\rm jet}$ is relatively weak, $\propto$ $\gamma_{\rm jet}^{-1/3}$, and the dependence on $\sigma$ is probably even weaker. For example, \citet{Chatterjee2019} found that adopting value for $\sigma_{\rm max}$ of 3, 10, 50, and 100 resulted in maximum Lorentz factors of $\approx$ 2, 5, 8, and 10, while \citet{McKinney2006} found that a value of $\sigma_{\rm max} = 10^4$ resulted in a maximum Lorentz factor of $\approx$ 10.  While both of these works assumed axisymmetry (thus not able to capture kink instabilities) and empty polar regions, these assumptions would, if anything, result in a higher conversion efficiency from magnetic to kinetic energy and a stronger dependence of $\gamma_{\rm jet}$ on $\sigma_{\rm max}$ than in our 3D, kink unstable jets that have to continually fight against the ambient medium.  This is encouraging but by no means conclusive evidence that $r_{\rm jet}$ and the evolution of the kink instability in our simulations are not strongly sensitive to the arbitrary choice of $\sigma_{\rm max}$.

\section{Comparison To Previous Work}
\label{sec:comp}
\subsection{Non-relativistic Simulations}
The reconnection-driven convection present in our simulations is a feature of both convection-dominated Bondi flow (CDBF) models \citep{CDBF} and magnetically frustrated models \citep{Pen2003,Pang2011}.  The fundamental difference between these two classes of solutions is the amount of energy transported by convection, $F_{\rm conv}$.  In CDBF models $F_{\rm conv}$ is large and positive, dominating the total energy transport budget and thus requiring $\rho \propto r^{-1/2}$ in steady state.  In the magnetically frustrated model, on the other hand, the magnitude of $F_{\rm conv}$ is relatively small and can even be negative (that is, carrying energy inwards). This allows for the power law index, $p$, of the radial density profile to be anywhere between $-3/2$ and $-1/2$ and relatively unconstrained otherwise (see \citealt{Gruzinov2001,Gruzinov2013} for discussions).  In essence, $F_{\rm conv}$ is suppressed because the magnetic field damps the convective motions, ``frustrating'' its ability to transport energy.   In a survey over initial magnetic field strength and Bondi radius, \citet{Pang2011} found $p \approx -1$ for low angular momentum flows of this type.

The convective flux in our simulations is weak and directed outwards (Figure \ref{fig:F_conv_comp}), consistent with the the fact that we find a $p\approx -1$ density profile that is steeper than the CDBF $p =-1/2 $. At certain times (e.g., top right and bottom left panel of Figure \ref{fig:Bx_contour}), the density and magnetic field distribution in our $\psi = 90^\circ$ simulation looks quite similar to those in the \citeauthor{Pen2003} (\citeyear{Pen2003}, their Figure 1) and \citeauthor{Pang2011} (\citeyear{Pang2011}, their Figure 1) simulations, with turbulent motions creating fairly random structure with no preferred direction.  At other times in the same simulation (e.g., bottom right panel of Figure \ref{fig:Bx_contour}) and at all times (after the initial transient period) in the other simulations, however, the jets powered by the spinning black hole evacuate the polar regions of matter and localize the turbulence to the midplane.    These black hole spin-powered jets have no analog in non-relativistic MHD.

While we do see a reduced inwards radial velocity compared to the hydrodynamic case (as evidenced by the critical sonic point moving inwards from $\sim$ 8.8 $r_{\rm g}$ to $\sim$ 2 $r_{\rm g}$ (Figure \ref{fig:radial_plots}), we do not find the $\sim$ zero radial velocities characteristic of the nearly hydrostatic CDBF and magnetically frustrated solutions.  This is a strictly GR effect, as all flows must pass through a sonic point before reaching the event horizon.  The stronger outwards force provided by a combination of magnetic fields and pressure can only delay the inevitable by moving the sonic point towards smaller radii.  If we were to extend our simulations to larger $r_{\rm B}$, the convective/turbulent zone would also extend to larger radii and could plausibly reach a nearly hydrostatic state in the region where GR effects are negligible.

\subsection{GRMHD Torus/Jet Simulations}

Two of our simulations ($\psi = 0^\circ$ and $\psi = 60^\circ$) appear to be magnetically arrested based on the traditional build-up and dissipation cycle seen in $\phi_{\rm BH}$ and the associated fluctuations in accretion rate caused by the brief repulsion of matter from near the black hole.  The other two simulations ($\psi = 30^\circ$ and $\psi = 90^\circ$) do not appear to be fully arrested since $\phi_{\rm BH}$ and $\dot M$ are much more stable with time, yet they are likely just under the threshold in magnetic flux to reach the arrested state.  Given that conclusion, it is instructive to compare our results to previous GRMHD MAD simulations (e.g., \citealt{Igumenshchev2003,Sasha2011,Narayan2012,White2019,Chael2019}).
To first approximation, Figures \ref{fig:Bz_contour}--\ref{fig:Bx_contour}, which show 2D slices of mass density and magnetic field lines, look remarkably similar to GRMHD MAD simulations at most times.  Both have evacuated, strongly magnetized polar regions with a large amount of coherent/vertical magnetic field lines.  Both see accretion via thin, non-axisymmetric streams due to the magnetic field strongly compressing the inflow close to the horizon (e.g., \citealt{Igumenshchev2003} ; see also, our Figure \ref{fig:beta_contour} which shows midplane slices of $\beta$ for $\psi = 30^\circ$).  These similarities between torus-based MADs and our initially zero angular momentum simulations are, perhaps, not that surprising because MAD flows are generally more sub-Keplerian than SANE flows (e.g., \citealt{Narayan2012}) and the magnetic fields dragged by the $a=0.9375$ black hole in our models provide enough torque on the gas to result in rotation rates that are a significant fraction of Keplerian near the horizon (top panel of Figure \ref{fig:radial_plots}).    

Jets in torus simulations without any other ambient medium are generally able to freely expand radially since the polar regions are essentially vacuum. This allows for them to reach much larger distances and Lorentz factors than we see in our simulations (e.g., \citealt{McKinney2006,Chatterjee2019}).  In this respect, our simulations are more similar to those performed by \citet{Bromberg2016,Sasha2016,Rodolfo2017}, where the jets were embedded in spherical interstellar mediums of various density profiles $\rho^{-s}$.  \citet{Bromberg2016} provided an analytical model for the jet as a function of height and found that for $s<2$ it becomes more unstable to the external kink instability at larger distances from the black hole, predicting that this would lead to the jet stalling or even breaking up.  Consistent with that expectation, the jets in our simulations (which show $s \approx 1$ for $r \lesssim r _{\rm Bondi}$ and $s \approx 0$ for $r\gtrsim r _{\rm Bondi}$) do both of those things (Figure \ref{fig:jet_radius} and \ref{fig:Bz_contour_jet}--\ref{fig:Bx_contour_jet}).  Note that the value of $s$ in this work for $r \lesssim r_{\rm B}$ is not a parameter but is determined by the dynamics of the accretion flow.  

Simulations of tilted disks find that the jets at large radii tend to align perpendicular to the large-scale disc (e.g., \citealt{White2019b,Liska2020}).   This is primarily caused by the jet propagating through the pre-evacuated, low-density funnel carved out by the initial torus. This region is the path of least resistance for the Poynting flux to travel outwards (as opposed to penetrating through the outer disc). Since our simulations start with a zero angular momentum, spherical distribution of gas, the jets have no other option but to penetrate through the ambient medium.  This prevents them from getting systematically channelled away from the black hole spin axis.

\subsection{Wind-Fed Accretion Onto Sgr A*}
We now compare our results to the simulations that inspired them in the first place, namely, the GRMHD wind-fed Sgr A* accretion simulations of \citet{Ressler2020b}, hereafter R20.  Both sets of simulations display magnetically confined midplanes, evacuated polar regions, magnetically arrested behavior, and relatively low $\beta$, coherent magnetic fields. Both also find $\rho$ $ \tilde \propto $ $r^{-1}$, though it is not clear that it is for the same reasons.  In \citet{Ressler2020b} the density profile was caused by the underlying distribution of angular momentum with accretion rate in material being fed from large radii, while here it is a result of `magnetically frustrated' convection in a pressure supported configuration.  Why exactly the latter results in an $r^{-1}$ density profile is an open question.  \citet{Gruzinov2013} argued that if there was some physical process enforcing a constant momentum flux (i.e., $\dot P \propto \rho v_r^2 r^2$ = const. in the non-relativistic regime) then $\rho \propto r^{-1}$ follows as a natural consequence.  $\dot P=$ const. can be achieved in a steady state only if gravity, pressure, and magnetic forces are perfectly balanced.  While this does seem to be the case in the non-relativistic simulations (e.g., Figure 2 in \citealt{Pen2003}), in our simulations $\dot P$ increases with decreasing radius.  The true reason for the $r^{-1}$ density scaling may just be the less satisfying observation that the simulations fall somewhere between Bondi-type flows and Convection Dominated Accretion Flow (CDAF)-type flows so the density power law index must fall somewhere in between $-3/2$ and $-1/2$. It is also not clear if our simulations have sufficiently large dynamic range in radius to determine the truly self-similar scaling of the density profile.

The most obvious difference between the accretion flow in this work and that found in R20 is the nature of the polar regions.  Since R20 used $a=0$,  the poles were evacuated due to a high concentration of vertical magnetic fields but lacked the toroidal field strength necessary to drive powerful jets.  As a result, the flow became quasi-spherical past a few $10$s of $r_{\rm g}$.  Furthermore, these magnetic poles changed direction with time (corresponding to variations in the net field being fed from larger radii), also tilting the gas away from the previous midplane.  As we have shown, because the black hole in our simulations is rapidly spinning, it strongly winds up the poloidal magnetic field in the toroidal direction, producing jets that reach to $\sim$ hundreds of $r_{\rm g}$.  While the jets at large radii can be significantly tilted with respect to the black hole spin due to propagation effects and the kink instability, near the horizon the they are always aligned with the spin axis regardless of initial magnetic field orientation.   We expect that if the R20 simulations were  run with a large black hole spin, they would undoubtably feature a jet that would disrupt the nature of the flow out to a relatively large radius and would most likely be less prone to oscillations in the location of the midplane.

\subsection{Spherical Infall Around Post-Merger Black Hole}
As this work was in the latter stages of preparation, \citeauthor{Kelly2020} (\citeyear{Kelly2020}, hereafter K20) appeared on arXiv presenting the results of a parallel study of low angular momentum GRMHD accretion onto rotating black holes in the context of black hole merger remnants, focussing mainly on the orientation and power of the resulting jet.  The initial conditions used in their simulations were essentially the same as those used here: uniform pressure, density, and magnetic field of varying inclination angles with respect to the black hole spin.  The main differences were that they 1) assumed a black hole spin of $a= 0.69$, lower than our $a=0.9375$, 2) smaller Bondi radii (higher initial gas temperatures), $r_{\rm B} \approx$ 8--81 $r_{\rm g}$ compared to our $r_{\rm B}=200$ $r_{\rm g}$, and 3) an adiabatic index appropriate for radiation pressure-supported gas, $\gamma=4/3$.   Their fiducial model had $r_{\rm B} \approx$ 13 $r_{\rm g}$

In contrast to our results, K20 found that the jets in their simulations only aligned with the spin axis of the black hole for $r \lesssim$ 30 $r_{\rm g}$, while at larger radii the jets aligned with the initial magnetic field direction.  Moreover, they found jet power to monotonically decrease with $\psi$ and found no evidence of the $\psi \sim$ 40--80$^\circ$ peak in $\eta$ as we do (see bottom panel of Figure \ref{fig:mdot_time}, Figure \ref{fig:phi_vs_psi}, and \S \ref{sec:60}) seen in our simulations.   We believe that this is predominantly because the K20 simulations are in an entirely different parameter regime than ours, where the dynamical range in radius is very small and any extrapolation in Bondi radius breaks down.  This can be seen from their plots of various quantities vs. `$\kappa$', their proxy for the entropy/temperature of the initial gas which relates directly to $r_{\rm B}$.  For $r_{\rm B} \lesssim 13$ $r_{\rm g}$, their simulations are roughly independent of $r_{\rm B}$, while for $r_{\rm B} \gtrsim$ 13 $r_{\rm g}$ they show an increase of $|\dot M|$, $\dot E$, and $\eta$ with increasing $r_{\rm B}$.  Moreover the simulations with smaller $r_{\rm B} $ are much less variable in time than those with larger $r_{\rm B}$. Thus, there is a clear transition at $r_{\rm B} $ $\sim$ 13 $r_{\rm g}$; the regime of $r_{\rm B} \lesssim$ 13 $r_{\rm g}$ is appropriate for some merger scenarios (and perhaps other stellar mass sized black holes), while the regime of $r_{\rm B} \gtrsim$ 13 $r_{\rm g}$ is appropriate for active galactic nuclei.  Since K20 focused on the former while we focus on the latter, our works are complementary.


\section{Conclusions}
\label{sec:conc}

We have presented GRMHD simulations of spherical accretion onto a rotating black hole modified by the presence of uniform, relatively weak magnetic fields of various tilt angles with respect to the black hole spin.  As the field is advected inwards it forms current sheets and reconnects,  dissipating magnetic energy and driving turbulence (Figures \ref{fig:Bz_contour}--\ref{fig:Bx_contour}).  Similar to previous non-relativistic `magnetically frustrated' models, we find that convection does not efficiently transport energy outwards (Figure \ref{fig:F_conv_comp}) 
and the mass density scales as $\tilde \propto$ $ r^{-1}$ (Figure \ref{fig:radial_plots}).  Our simulations are either magnetically arrested or very close to magnetically arrested, with the dimensionless magnetic flux, $\phi_{\rm BH}$ saturating at 20--80, often showing limit cycles of slow build-up followed by rapid dissipation (Figure \ref{fig:mdot_time}).  Torque caused by frame dragging of magnetic field lines threading the rapidly spinning black hole is enough to result in significant (up to $\sim$ 0.5 Keplerian) angular velocities near the event horizon (Figure \ref{fig:radial_plots}) which may be enough to cause Doppler asymmetry in the 230 GHz images (e.g., as in EHT observations of M87, \citealt{EHT1}).   

All of our simulations form strongly magnetized jets that predominantly align with the spin axis of the black hole.  While there are times when the jets can be significantly tilted, this tilt does not seem to correlate with $\psi$ (the initial magnetic field direction), but rather results from symmetry breaking as the turbulently fed jet plows through the surrounding gas (Figures \ref{fig:Bz_contour_jet}--\ref{fig:Bx_contour_jet}).  At larger radii the jet dynamics become dominated by the kink instability (Figure \ref{fig:jet_spacetime}), ultimately becoming disrupted at a few hundred $r_{\rm g}$ (Figure \ref{fig:jet_radius}) at which point they transition into an uncollimated outflow of hot gas reaching to $\sim$ 1000 $r_{\rm g}$ (Figure \ref{fig:jet_velocity}). The dissipation mechanism of the kink modes is likely magnetic reconnection near the jet wall boundary \citep{Bromberg2019}, which could in turn accelerate particles to high energy \citep{Davelaar2020} and contribute to nonthermal emission.  


The nature of our simulations is highly stochastic.  Not only are the dynamics of the accretion flow governed by reconnection-driven turbulence, the dynamics and shape of the jet are governed by the kink instability and interaction with the turbulent ambient medium.  This is evident by the various contours of density, $\beta$, and $b^2/\rho$ plotted in Figures \ref{fig:Bz_contour}--\ref{fig:Bx_contour}, \ref{fig:beta_contour}, and \ref{fig:Bz_contour_jet}--\ref{fig:Bx_contour_jet} which vary dramatically as a function of time and frequently display large global asymmetries.   Small changes in the initial conditions such as the tilt of the magnetic field ($\psi$) could therefore lead to completely different realizations of the flow, accounting for much of the variation between the simulations we have presented.  All things considered, the dynamics of the flow do not depend strongly on $\psi$ except for a couple of special cases.  

The first case is $\psi = 90^\circ$ where there is exactly zero net initial vertical flux (and presumably for $\psi$ close to 90$^\circ$):  the jet takes longer to form, is less efficient than the other simulations, and is intermittent even at horizon scales; all because vertical flux can only be generated through turbulent motions.  The second case is $\psi \sim$ $40^\circ$--80$^\circ$ where there is a peak in the magnetic flux threading the horizon, $\phi_{\rm BH}$, (Figure \ref{fig:phi_vs_psi}) that is a factor of $\lesssim$ 2 higher than simulations with $\psi\lesssim$ $40^\circ$ and $\psi \gtrsim$ $80^\circ$.  We suspect that this range in angles represents a ``sweet spot'' where there is slightly less turbulence inhibiting the accretion of magnetic flux because the large-scale current sheet is partially blown away by the polar outflow and yet there still enough initial vertical flux to form a jet before the onset of reconnection (see \S \ref{sec:60}).

The fact that even the $\psi = 90^\circ$ simulation forms (at times) a relatively high efficiency jet with $\eta \lesssim 20 \%$ (third panel of Figure \ref{fig:mdot_time}) gives further evidence that a large-scale net vertical flux is not necessarily a prerequisite for \citet{BZ1977} jet formation.  This was previously shown for flows with larger angular momentum where an MRI-driven dynamo can convert toroidal magnetic field into poloidal magnetic field \citep{Parfey2015,Christie2019,LiskaDynamo}.  Here we find that turbulent motions alone can produce enough local net vertical magnetic flux to power a jet from an already mostly poloidal but zero net vertical flux field.  These results suggest that jets may be ubiquitous in systems with rapidly rotating black holes regardless of the magnetic field geometry in the accretion flow's source.

The simulations presented in this work may be a reasonable model for Sgr A*.  Magnetically arrested-type flows have been favored in several recent analyses comparing models to observations due to 1) the relatively high levels of linear polarization observed in the mm emission (models with less magnetic flux tend to require higher accretion rates that can depolarize the emission, \citealt{JR2018,Dexter2020}) and 2) the relatively large amount of vertical magnetic field required to reproduce the NIR flare polarization periodicity \citep{GRAVITYFlare,Dexter2020b,Porth2020,Gravity2020}.    Moreover, wind-fed GRMHD accretion simulations that take into account the WR stellar winds primarily sourcing the accretion flow naturally evolve into an arrested state \citep{Ressler2020b}.  On the other hand, magnetically arrested models are known to produce powerful jets (at least for rapidly spinning black holes, e.g., \citealt{Sasha2011}) and yet there is no unambiguous detection of a collimated jet in Sgr A*.  Our simulations suggest, however, that this fact does not necessarily rule out the arrested model.   Feeding by stellar winds can result in a quasi-spherical distribution of gas outside of the inner horizon-scale that pressure-confines the jet prompting dissipation through the kink instability.  If this happens well inside the Bondi radius then larger scale observations may only be able to probe the hot, matter-dominated outflow blown out by the unstable jet `head' as it dissipates (e..g, Figure \ref{fig:jet_velocity}). In fact, recent work using ALMA may have detected just that, finding evidence for a high-velocity, bipolar outflow with an opening angle of $\sim$ 30$^\circ$ extending from $\sim$ 7$\times 10^5$ $r_{\rm g}$ to $\sim$ $10^8$ $r_{\rm g}$  \citep{Royster2019,YZ2020}.
Alternatively, Sgr A* could be slowly spinning or the jet may be pointed directly at us.
Finally, the $r^{-1}$ dependence of the mass density that we find in our simulations (Figure \ref{fig:radial_plots}) is consistent with observations that use different methods to probe multiple different scales of the gas surrounding Sgr A* \citep{Xu2006,Gillessen2019}.   

It is less clear whether our simulations represent a viable model for M87 or other similar galactic nuclei.   
Resolved X-ray emission at the Bondi radius of M87 can be fit with an $\rho \propto r^{-1}$ distribution \citep{Russell2015}, as in our simulations, though it is unknown if this dependence persists to smaller radii.  
In order to reproduce the unambiguous, well-collimated jet observed out to$\sim$ $10^6$--$10^7$ $r_{\rm g}$ we would require the jets in our simulations to be stable to kink instabilities out to radii $\gtrsim $ $10^4$ times what we see here (Figure \ref{fig:jet_radius}).   
Simply increasing the Bondi radius to a more realistic value may accomplish this, but based on an (admittedly uncertain) extrapolation we have argued that the radius at which the jet disrupts may be independent of Bondi radius (see \S \ref{sec:bondi_ext} and the next paragraph for a discussion). 
This could indicate that M87 is fed in a different manner than the Galactic Centre, with less low angular momentum gas and more disk-like accretion.  
Indeed, even if the M87 black hole is predominantly fed by stellar winds (which is not clear given the larger hot gas reservoir in the inner intracluster medium of Virgo), the nature of this feeding \emph{must} be different in some respects since the much larger (absolute) accretion rate requires a much larger number of winds to sustain.  This is especially true given that in Sgr A* only a handful of winds end up contributing to the accretion rate \citep{Loeb2004,Cuadra2008,Ressler2018}.

The self-consistent formation of the rising, buoyant bubbles (or cavities) in our simulations due to a combination of the kink instability and time variability of the jet (Figures \ref{fig:Bz_contour_jet}--\ref{fig:Bx_contour_jet}) may be broadly analogous to what occurs in galaxy clusters (e.g., \citealt{Fabian2000,Blanton2010}), though the spatial and temporal scales we simulate are orders of magnitude smaller than those observed in clusters.  If the estimated scalings of our results with $r_{\rm B}$ (\S \ref{sec:bondi_ext}) are correct, then even for much larger $r_{\rm B}$ these cavities would never reach the sizes and distances required to match observations.  On the other hand, if the flow had more angular momentum so that the poles were pre-evacuated, the jet could potentially reach the required distances before becoming kink unstable enough to produce bubbles.

For computational reasons we have focused on one particular Bondi radius ($r_{\rm B} = 200 r_{\rm G}$) that is orders of magnitude less than the actual Bondi radii for real systems, including both M87 and Sgr A*.  Using the dependence on radius of $\rho$ and the kink stability criteria, we extrapolated to larger $r_{\rm B} \gg 200 r_{\rm g}$ (\S \ref{sec:bondi_ext}) to predict that $\dot M/\dot M_{\rm Bondi}$  $\propto$ $r_{\rm B}^{-1/2}$ and $r_{\rm jet}$  $\propto$ const, where $r_{\rm jet}$ is the radius at which the jet disrupts by the kink instability.  However, since we have not rigorously tested these dependencies by running additional simulations with larger $r_{\rm B}$,  they should be taken with a grain of salt.   Future work can test these predictions using a parameter survey in $r_{\rm B}$ or by using the actual Bondi radius in a multi-scale, multiple simulation technique to extend the dynamic range in the manner of R20.

We have also focused on one particular value of $\sigma_{\rm max}$, the maximum $\sigma \equiv b^2/\rho$ allowed in the simulations, as well as one particular value of the initial $\beta$.  It is unclear the degree to which our results are affected by these choices.  In \S \ref{sec:beta} and \S \ref{sec:sigma} we argue that the dependence should be relatively weak based on previous simulations in the literature, though none of those works had an entirely similar set-up as we have here. For $\beta$, the hope is that for any initial $\beta \gg 1$, the inwards advection of magnetic field would ultimately lead to approximately the same steady state with larger $\beta$ just taking longer to reach that state.  For $\sigma_{\rm max}$, realistic astrophysical jets likely have $\sigma_{\rm max}$ orders of magnitude larger than the $\sigma_{\rm max}=100$ imposed here, which if the conversion of magnetic to kinetic energy in the jets was 100\% efficient, this would lead to orders of magnitude larger Lorentz factors. This is an important consideration since jets with larger Lorentz factors could be even more unstable to kink modes (Equation \ref{eq:kink}).   However, GRMHD jets tend to be rather inefficient at converting magnetic to kinetic energy even in 2D where they are kink stable,  which makes $\gamma_{\rm jet}$ not as strongly dependent on $\sigma_{\rm max}$.  We expect this to be even more true in the 3D, kink unstable jets we have presented here and thus we can reasonably hope that the main conclusions drawn in this work are not strongly sensitive to $\sigma_{\rm max}$.

\section*{Acknowledgments}
We thank S. Philippov for useful discussions, J. Stone and C.~F. Gammie for comments on the manuscript, as well as all the members of the Horizon Collaboration, \href{http://horizon.astro.illinois.edu}{http://horizon.astro.illinois.edu}. SMR was supported by the Gordon and Betty Moore Foundation through Grant GBMF7392.  SMR also thanks R. and D. Ressler for
their generous hospitality during the writing of this manuscript.
This work was supported in part by NSF grants NSF PHY--1748958,  AST--1715054, AST--1715277, a Simons Investigator award from the Simons Foundation, and by the NSF
through XSEDE computational time allocations TG--AST170012 and TG--AST200005 on Stampede2.  This work was made possible by computing time granted by UCB on the Savio cluster.

\section*{Data Availability}
The data underlying this paper will be shared on reasonable request
to the corresponding author.

\bibliographystyle{mn2efix}
\bibliography{mmsa}

\end{document}